\documentclass[prb, twocolumn, showpacs, superscriptaddress,longbibliography]{revtex4-1}
\usepackage{amsmath, amsfonts, amssymb, mathrsfs}
\usepackage{graphicx, subfigure, color}
\usepackage{bm}
\usepackage{braket}
\usepackage{epstopdf}
\usepackage{color}
\usepackage[breaklinks]{hyperref}
\hypersetup{
   colorlinks,
   citecolor=blue,
   filecolor=blue,
   linkcolor=blue,
   urlcolor=blue
}
\graphicspath{Graphes}

\newcommand{\be}{\begin{equation}}
\newcommand{\ee}{\end{equation}}

\definecolor{grey}{rgb}{.6,.6,.6}

\begin{document}

\title{Phase Diagram and Entanglement of two interacting topological Kitaev chains}

\author{Lo\"{i}c Herviou}
\affiliation{Centre de Physique Th\'{e}orique, \'{E}cole Polytechnique, CNRS, Universit\' e Paris-Saclay, 91128 Palaiseau, France}
\affiliation{Laboratoire Pierre Aigrain, \'Ecole Normale Sup\'erieure-PSL Research University, CNRS, Universit\'e Pierre et Marie Curie-Sorbonne Universit\'es, Universit\'e Paris Diderot-Sorbonne Paris Cit\'e, 24 rue Lhomond, 75231 Paris Cedex 05, France}
\author{Christophe Mora}
\affiliation{Laboratoire Pierre Aigrain, \'Ecole Normale Sup\'erieure-PSL Research University, CNRS, Universit\'e Pierre et Marie Curie-Sorbonne Universit\'es, Universit\'e Paris Diderot-Sorbonne Paris Cit\'e, 24 rue Lhomond, 75231 Paris Cedex 05, France} 
\author{Karyn~Le~Hur}
\affiliation{Centre de Physique Th\'{e}orique, \'{E}cole Polytechnique, CNRS, Universit\' e Paris-Saclay, 91128 Palaiseau, France}

\begin{abstract}
A superconducting wire described by a p-wave pairing and a Kitaev Hamiltonian exhibits Majorana fermions at its edges and is topologically protected by symmetry. We consider two Kitaev wires (chains) coupled by a Coulomb type interaction and study the complete phase diagram using analytical and numerical techniques.  A topological superconducting phase with four Majorana fermions occurs until moderate interactions between chains. For large interactions, both repulsive and attractive, by analogy with the Hubbard model, we identify Mott phases with Ising type magnetic order. For repulsive interactions, the Ising antiferromagnetic order favors the occurrence of orbital currents spontaneously breaking time-reversal symmetry. By strongly varying the chemical potentials of the two chains, quantum phase transitions towards fully polarized (empty or full) fermionic chains occur. In the Kitaev model, the quantum critical point separating the topological superconducting phase and the polarized phase belongs to the universality class of the critical Ising model in two dimensions. When increasing the Coulomb interaction between chains, then we identify an additional phase corresponding to two critical Ising theories (or two chains of Majorana fermions). We confirm the existence of such a phase from exact mappings and from the concept of bipartite fluctuations. We show the existence of negative logarithmic corrections in the bipartite fluctuations, as a reminiscence of the quantum critical point in the Kitaev model. Other entanglement probes such as bipartite entropy and entanglement spectrum are also used to characterize the phase diagram. The limit of large interactions can be reached in an equivalent setup of ultra-cold atoms and Josephson junctions.
\end{abstract}

\date{\today}
\maketitle

\section{Introduction} 
The discovery of topological effects in  quantum solid states has revolutionized condensed matter studies. From the  fractional Quantum Hall effect to topological superconductors, topological Hamiltonians have been a topical subject of study during the last few years. One of the main characteristics of such systems is the presence of anionic excitations, with fractional spin and charge. In particular, in the case of topological superconductors, zero-energy Majorana modes appear at the edges of these systems or in vortices\cite{Read2000,Kitaev2001}. It is of particular interest in conjunction with the rise of quantum information\cite{Bert}. The topological nature of these modes, preventing any effect from small local interactions, protects them from decoherence and make them perfect candidates for the realization of quantum bits\cite{Nayak2008}. Indeed, several schemes have been proposed to realize complete sets of quantum gates and memories, using superconducting wires with Majorana fermions at each extremities \cite{Beenakker2013, Hyart2013, Aasen2015}. These schemes rely on controlled interactions between several of such wires. The natural question concerns the effects of other uncontrolled interactions that could arise due to the proximity of these wires. Lately, numerous propositions on topological systems presenting solvable points have been made\cite{Iemini2015, Buechler,Katsura2015}. We present in this paper the general study of two topological superconducting wires in the presence of the simplest interaction, a Coulomb-like interaction modeled by an on-site repulsion \`a la Hubbard. It is a generic model, in the sense that these interactions will be present in most systems. This ladder is also a building step towards building two-dimensional materials.\\
The simplest model of superconducting wire is the well-known and exactly solvable Kitaev's wire\cite{Kitaev2001} :
\begin{multline}
H_K\{c\}=-\mu \sum\limits_{j=1}^{L} c^\dagger_{j}c_{j} + \sum\limits_{j=1}^{L-1} -t( c^\dagger_{j} c_{j+1} + c^\dagger_{j+1} c_{j})\\
+\Delta (c_{j}^\dagger c_{j+1}^\dagger+  c_{j+1} c_{j}).
\end{multline}
Here, $c$ is a fermionic spinless annihilation operator, $t$ is the hopping amplitude (it will serve as an energy scale in the rest of the paper), $j$ is a site index and $\Delta$ is a Bardeen-Cooper-Schrieffer (BCS) \cite{BCS} p-wave superconducting pairing term generated by an interaction with a superconducting substrate. $L$ is the number of sites in the wire. There has been several proposals and  realizations of this model, for example by coupling a semiconducting nanowire to the bulk of two- or three-dimensional superconductors via a strong spin-orbit interaction and by applying a magnetic field to select one spin species in the wire\cite{wire1,wire2,V.Mourik2012,A.Das2012,Aasen2015}.  Other implementations have been discussed with ferromagnetic metallic chains \cite{Ali2,Ali} and ultra-cold atoms \cite{Bardyn, Kraus2013}. Majorana fermions can also occur as a result of purely intrinsic attractive interactions \cite{Berg}. This model presents a $\mathbb{Z}_2$ topological degeneracy in its ground state, corresponding to a free Majorana fermion subsisting at each extremity of the wire. 

We consider two interacting Kitaev chains, coupled via a Coulomb interaction:
\begin{equation}\label{eq:inter}
H_{\text{int}}=g \sum\limits_j (n_{j,1}-\frac{1}{2})(n_{j,2}-\frac{1}{2}) ,
\end{equation}
$n_{j,1/2}$ is the electron number operator in the first/second wire at site $j$. Interpreting the chain index as a spin index, then this can be identified as the well-known Hubbard interaction, a staple of condensed matter physics thoroughly studied for the last 40 years. This interaction does not break any of the discrete symmetries of the original problem. Indeed, while Kitaev's model does not conserve the number of fermions, it preserves fermionic parity and has time-reversal symmetry (and particle-hole symmetry for zero chemical potential).  Introducing a second wire, though, has non trivial effect on the topology: following the classification of topological phases proposed by Fidkowski and Kitaev~\cite{Fidkowski2011,Fidkowski2010}, we know that the topological nature of a set of $n$ identical wires in interactions depends on $n$. In particular, with 2 wires, we know that the system is only a symmetry protected topological phase (SPT). Hence, an arbitrarily small term breaking one of the original symmetries of the model, in our case time reversal symmetry, lifts the degeneracy of the ground state. 

A large variety of interactions can be considered. In particular, fine-tuned interacting terms have been added to make the Kitaev ladder exactly solvable \cite{Iemini2015,Buechler}. Supplementary terms could be considered such as introducing a hopping term between the two wires $-t_\perp c^\dagger_1 c_2 + h.c$ or allowing for an orthogonal pairing term $\Delta_\perp c^\dagger_1 c^\dagger_2 + h.c$ \cite{Potter2010}. While these terms can be tuned to obtain exactly solvable points, they also become negligible for a large enough distance $d$ between the wires.  The Coulomb repulsion scales like $1/d^2$. The hopping amplitude, scaling as $\exp (-d/\chi)$, with $\chi$ being a correlation length, is negligible for $d\gg \chi$. Similarly, if $d$ is larger than the coherence length of the Cooper pair, one can safely ignore $\Delta_\perp$, as long as both these terms do not break the time-reversal symmetry, i.e $t_\perp$ and $\Delta_\perp$ are real. Several interacting terms have been also considered in the case of one wire \cite{Stoudenmire2011, Gangadharaiah2011, sela2011, Thomale2013, Chan2015}.  In this work, we ignore the effect of intra-wire repulsive interactions and assume that the Cooper channel dominates in each wire. This gives a minimal model, which interpolates between Hubbard and Kitaev physics, and displays a competition between topological superconducting ordering and Mott ordering. Reaching the large $g$ limit could be eventually achieved experimentally by placing an insulating material between the wires, forming a capacitance between the two parallel wires. Coupling with a bath could also allow to engineer such an interaction term (more complex interaction terms have been envisioned recently in Refs. \onlinecite{Iemini2015,Buechler}). Not that two side gates could be used to screen out the interactions along the two wires. Other interaction effects through charging energy terms could also produce topological Kondo boxes \cite{Beri2012,beri2013,Altland,Erik,galpin2014}. We also note that since a Kitaev superconducting wire can be engineered in ultra-cold atoms through proximity effect \cite{LeHur2001} or via a Floquet type approach \cite{Potter2016}, then a controllable interaction could be achieved between and inside the two wires.

This problem can also be mapped in terms of two interacting Ising spin-$\frac{1}{2}$ chains \cite{Subirbook}:
\begin{multline}
H_\text{spin}=\sum\limits_{w=1}^2\left(- \sum\limits_{j=1}^L\mu \sigma_{j,w}^z +\sum\limits_{j=1}^{L-1}\frac{(\Delta-t)}{2} \sigma^x_{j,w} \sigma_{j+1,w}^x  \right.\\
\left. -\frac{(\Delta+t)}{2} \sigma^y_{j,w} \sigma_{j+1,w}^y  \right)+g\sum\limits_{j=1}^L \sigma_{j,1}^z \sigma_{j,2}^z,
\end{multline}
where $w$ is a chain index and $\sigma^{x,y,z}$ the Pauli matrices. This model will have the same phase diagram as its fermionic counterpart,  but different physical properties \citep{Greiter2014}. This representation favors another controlled experimental realization with cold atoms \cite{Zollerpipeline, simon2011quantum} or using Josephson junctions as pseudo two-level systems \cite{Levitov2001}, allowing to access the large $g$ limit. Other Majorana-Josephson models, similar to Ising models in transverse fields have been proposed, see for example Ref. \onlinecite{Terhal2012}. Such systems would allow us to reach the high coupling limits, and consequently to probe easily the more exotic features of our system. Quantum criticality in a Ising chain has also been observed in real materials \cite{Coldea2010}.

\begin{figure}
\begin{center}
\includegraphics[width=0.5\textwidth]{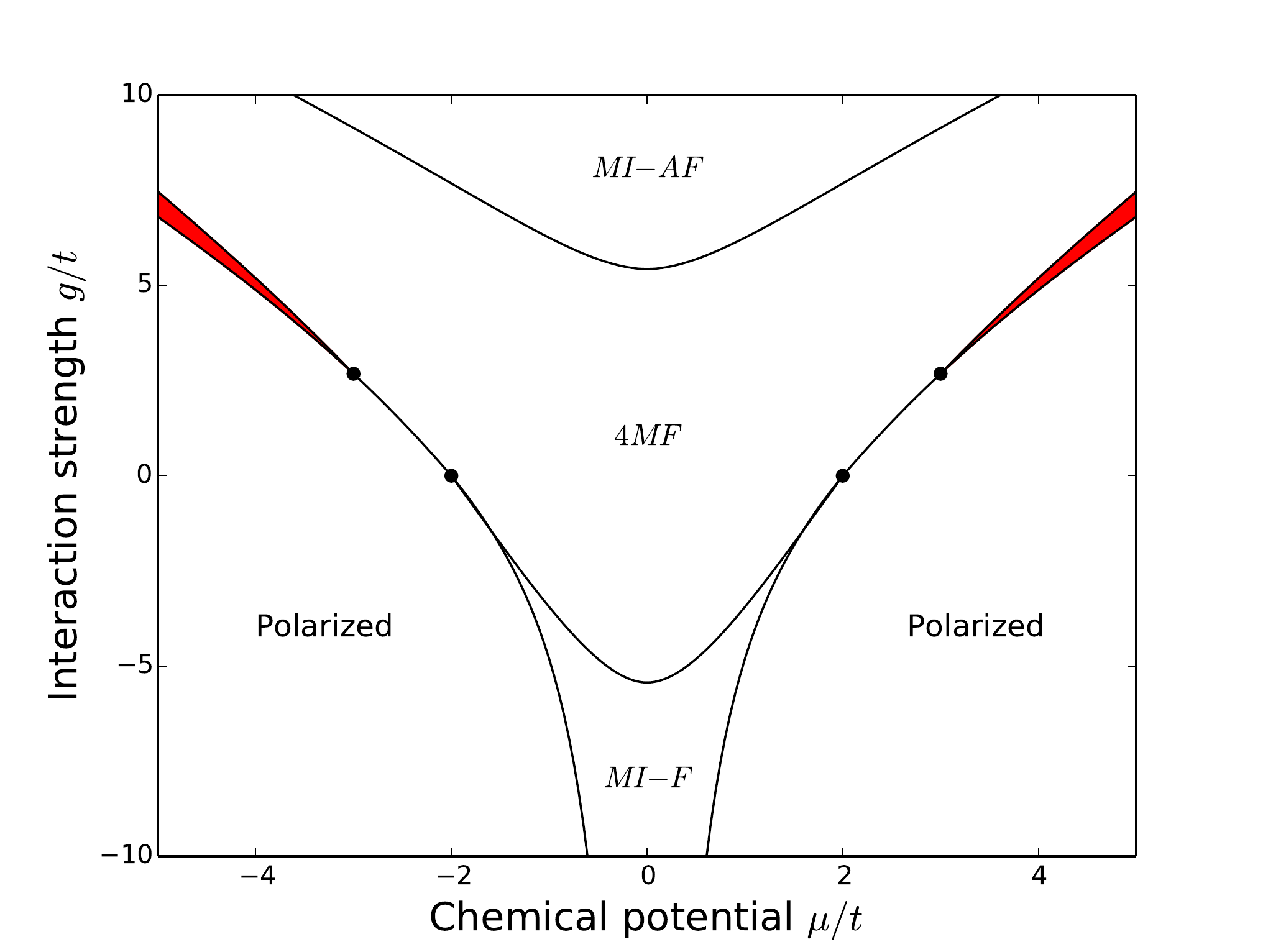}
\caption{(Color online) Sketch of the phase diagram of the interacting ladder at $\Delta=t$ obtained with analytical (bosonization, exact mappings) and numerical methods (exact diagonalization (ED), Density Matrix Renormalization Group (DMRG)). The chemical potentials of the two chains are taken to be equal. $4MF$ is the $SPT$ gapped phase presenting 2 Majorana fermions at each of the ladder extremities. $MI-AF$ and $MI-F$ are two gapped Mott phases, either antiferromagnetic or ferromagnetic. Polarized corresponds to a trivial phase with a quasi-empty or quasi-full ladder. In red, the gapless $DCI$ phase embodies an extension of the critical point  at $g=0$. It acquires an extension of order $t$ as $g$ goes to  $+\infty$.}
\label{fig:CompletePD}
\end{center}
\end{figure}

The main result of this paper is the phase diagram of our model for $\Delta \neq 0$ in Fig. \ref{fig:CompletePD}. We observe the survival of the SPT phase, called $4MF$, in the presence of finite interactions. This phase is characterized by  two free Majorana fermions at each extremity of the ladder. Despite their proximity, the absence of an appropriate pairing or hopping term between the two ladders prevents a direct coupling between the Majorana fermions at each extremity. We consequently observe a four-fold degeneracy of the ground state of the system with open boundary conditions. Each of these ground states has a different combination of fermionic parities. At very large coupling and weak chemical potential, two similar phases appear. Both of them are Mott-Ising phases related to the Mott phases of the Hubbard model. For positive $g$ ($MI$-$AF$), the corresponding low-energy model is an Antiferromagnetic Ising model, which presents orbital currents and a spontaneous breaking of the time reversal symmetry. For $g$ negative ($MI$-$F$), the low-energy model is a Ferromagnetic Ising model in a transverse field, which also breaks time reversal symmetry and exhibits currents between the two wires. At large chemical potential, a polarized trivial phase opens, corresponding to depleted or full wires. At finite positive coupling, an intermediate phase opens between $4MF$ and the polarized phase. This is the only gapless phase in this diagram and has a central charge $c=1$. This phase is an extension of the critical point at $g=0$, whose critical model is two Ising models. We will denote it Double Critical Ising ($DCI$). It is nothing but a Luttinger Liquid (LL) of a complex mixture of fermions on the two wires.

Finally, it should be noted that other coupling forms in the two-chain model of spinless fermions give distinct phase diagrams \cite{urskaryn,Carr2006,Julia,Guillaume}. 

The outline of the paper is as follows. In Section II, we will present some well-known results on both Kitaev's wire and the Hubbard model, two solvable limiting cases of our system, and introduce the notations that will be used in the rest of the paper. We also present the results of some standard mean-field computations. It provides some insight on the opening of the $DCI$ phase at a critical interaction strength. The next Sections will be devoted to the properties of the phase diagram and the $DCI$ phase. In Section III, we detail an entanglement probe, the bipartite charge fluctuations \cite{song2012bipartite}, that we use for characterizing the $DCI$ phase.  After reminding some previously obtained results on bipartite fluctuations in Luttinger Liquids \cite{song2011entanglement}, we extend them to a more general case and compute its behavior for the critical Ising model. Using exact and numerical computations and Conformal Field Theory (CFT) arguments, we find that the critical Ising phase is characterized by a negative subdominant logarithmic contribution to the bipartite fluctuations. In Section IV, we use bosonization\cite{QP1DGiamarchi2004} to derive the phase diagram close to half-filling, and explicit the properties of the $4FM$ and the two $MI$ phases. In Section V, we develop effective models to describe the emergence of the $DCI$ phase, and present numerical confirmations of its existence. We finally conclude in Sec.~\ref{sec:conclusion}. Appendices are devoted to the details of computations.

\section{Limiting cases}

We discuss in this Section two limiting cases of our system, the Kitaev model in Sec.\ref{sec-kitaev} and the Hubbard model in Sec.\ref{sec-hubbard}. Then, we present a pedagogical mean-field argument for the existence of the $DCI$ in Sec. \ref{sec-meanfield}.

\subsection{Kitaev's wire}\label{sec-kitaev}
\subsubsection{Topological phases and Majorana fermions}
In the absence of interactions $U=0$, our system reduces to two uncoupled Kitaev wires \cite{Kitaev2001}. This model is one of the simplest models presenting a topological phase. In this Section, we consider a single chain and recall some of the main results for the sake of comparison and to introduce notations that will be used in the rest of the paper. 

The Kitaev Hamiltonian for the fermionic species $c$, with open boundaries conditions (OBC), is written as:
\begin{multline}
H_K\{c\}=-\mu \sum\limits_{j=1}^{L} c^\dagger_{j}c_{j} + \sum\limits_{i=1}^{L-1} -t( c^\dagger_{j,} c_{j+1} + c^\dagger_{j+1} c_{j})\\
+\Delta c_{j}^\dagger c_{j+1}^\dagger+ \Delta^* c_{j+1} c_{j}.
\end{multline}

The gauge invariance of the fermions allows for a cancellation of the phase of $\Delta$: let $\Delta=|\Delta |e^{i \phi}$, the application $c_j \rightarrow e^{i\frac{\phi}{2}}c_j$ transforms $\Delta$ in $|\Delta|$. Consequently, in the rest of this paper, we consider $\Delta$ real and positive. Nonetheless, it comes at the price of the loss of this gauge freedom, equivalent to the conservation of charge. We introduce the Majorana operators $\alpha_j$ by splitting the on-site fermions in their real and imaginary parts:
\begin{equation*}
c_{j} = \frac{1}{2}(\alpha_{2j+1} + i \alpha_{2j}).
\end{equation*}
In order for the fermions to respect their standard algebra, the Majorana fermions verify the well-known Clifford algebra:
\begin{equation*}
\{ \alpha_{j}, \alpha_{k} \} = 2\delta_{j,k}.
\end{equation*}
Placing ourselves at $t=\Delta$ to keep expressions simple, and focus on $\mu \leq 0$, we obtain:
\begin{equation}
H_K=-\frac{i \mu}{2}\sum\limits_{j}\alpha_{2j+1} \alpha_{2j}-it \sum\limits_{j} \alpha_{2j+1} \alpha_{2j+2}
\end{equation}
At the point $\mu=-2t$, it reduces to a simple chain of free Majorana fermions of size $2 L$, $H_K=-it \sum_{j=1}^{2 L} \alpha_{j} \alpha_{j+1}$, where $L$ is the original number of fermionic sites. This model is conformally invariant with a central charge $c=\frac{1}{2}$~\protect\cite{rahmani2015}.

We pose $\delta \mu = \mu + 2t$. The Kitaev's model can be rewritten in terms of the $\alpha$ operators.
\begin{equation}
H_K=-\frac{i \delta \mu}{2}\sum\limits_{j} \alpha_{2j+1} \alpha_{2j}-it \sum\limits_{j} \alpha_{j} \alpha_{j+1}.\label{Eq:MajoranaWire}
\end{equation}
Thus $\delta \mu$ favors the pairing of neighboring Majorana fermions every two sites and Majorana dimerization in the ground state. The sign of $\delta \mu$ differentiates two pairing, translated by one site, corresponding to the standard and topological superconducting phases. When $\delta \mu < 0$, the chemical potential dominates and couples Majoranas on the same site. The fermionic excitations in this case correspond to the physical fermions. In contrast to that, when $\delta \mu > 0$, the hopping dominates and couples Majoranas on neighbouring sites.There is a Majorana dangling uncoupled at each extremity of the wire ($\alpha_2$ and $\alpha_{2L+1}$ for the limiting case $\mu=0$). The phase transition occurs at $\mu=\pm 2t$.\\
In the topological phase, the ground state is two-fold degenerate due to these two zero modes. No local operator can distinguish the two degenerate ground states of the topological phase, while they have opposite fermionic parity. The corresponding operator is defined as:
\begin{equation}
P=\exp (i \pi \sum\limits_{j=1}^L c^\dagger_j c_j) = \prod\limits_{j=1}^L (1-2c^\dagger_j c_j) =\prod\limits_{j=1}^L i \alpha_{2j} \alpha_{2j+1}.
\end{equation}
The two ground states of the topological phase have opposite fermionic parity due to the following commutation rules:
\begin{equation}
\{\alpha_j^A, P\}=0~~~~~~[P, H]=0.
\end{equation}
These results can be also obtained by solving the Hamiltonian with periodic boundary conditions (PBC), using a Bogoliubov transform. Details can be found in Appendix \ref{app:Bogo}.

\subsubsection{Symmetries}

The Kitaev's model does not present any continuous symmetry. Indeed, the unusual pairing term breaks the conservation of the charge (number of fermions) down to the conservation of fermionic parity. From Equation \eqref{Eq:BCSGS}, one can easily see that, in the case of PBC, the ground state is odd ($P=-1$) in the topological phase and even ($P=1$) in the two trivial phases~\protect\cite{ortiz2014}. With OBC, the two ground states have different fermionic parities. With two non-interacting wires, denoting $P_1$ and $P_2$ the fermionic parities in each wire, similar arguments prove that the ground state is odd-odd in the topological phase ($P_1=P_2=-1$) and even-even in the trivial phase with PBC ($P_1=P_2=1$). Similarly with OBC, the four ground states correspond to ($P_1=\pm 1$, $P_2=\pm 1$).\\
There are two other discrete symmetries. The first one is the particle hole symmetry $c_j \rightarrow (-1)^j c^\dagger_j$ occurring at $\mu=0$. This implies the symmetry of the phase diagram for $\mu \rightarrow -\mu$, and half-filling at $\mu=0$. The second and most important one is the time-reversal anti-unitary symmetry, where $\mathcal{K}$ is complex conjugation. Spinless real-space fermions are left invariant by $T$ with our choice of gauge. As mentioned in the introduction, as long as local interaction terms are invariant under this symmetry (and fermionic parity), the topological phase is preserved.

\subsubsection{Link with quantum Ising model}
Let $\sigma^{x,y,z}$ be the Pauli matrices. Kitaev wire can directly be mapped on a spin model through the canonical Jordan-Wigner transform \cite{Subirbook}:
\begin{equation*}
\sigma_j^z = c^\dagger_j c_j - \frac{1}{2}~~~~\sigma^x_j=\frac{1}{2}(c_j+c^\dagger_j) \exp(i\pi \sum\limits_{l=1}^{j-1} c_l^\dagger c_l).
\end{equation*}
The obtained Hamiltonian can be written as:
\begin{equation}
H_I= \sum\limits_{j=1}^{L-1}\frac{(\Delta-t)}{2} \sigma^x_j \sigma_{j+1}^x -\frac{(\Delta+t)}{2} \sigma^y_j \sigma_{j+1}^y  - \sum\limits_{j=1}^L\mu \sigma_j^z.
\end{equation}

At $\Delta=t$, we recover the well-known Quantum Ising Model in a transverse field. Properties do not change when we leave this special point, as long as $\Delta, t \neq 0$.

\subsection{The Hubbard model}\label{sec-hubbard}

\begin{figure}
\begin{center}
\includegraphics[width=0.45\textwidth]{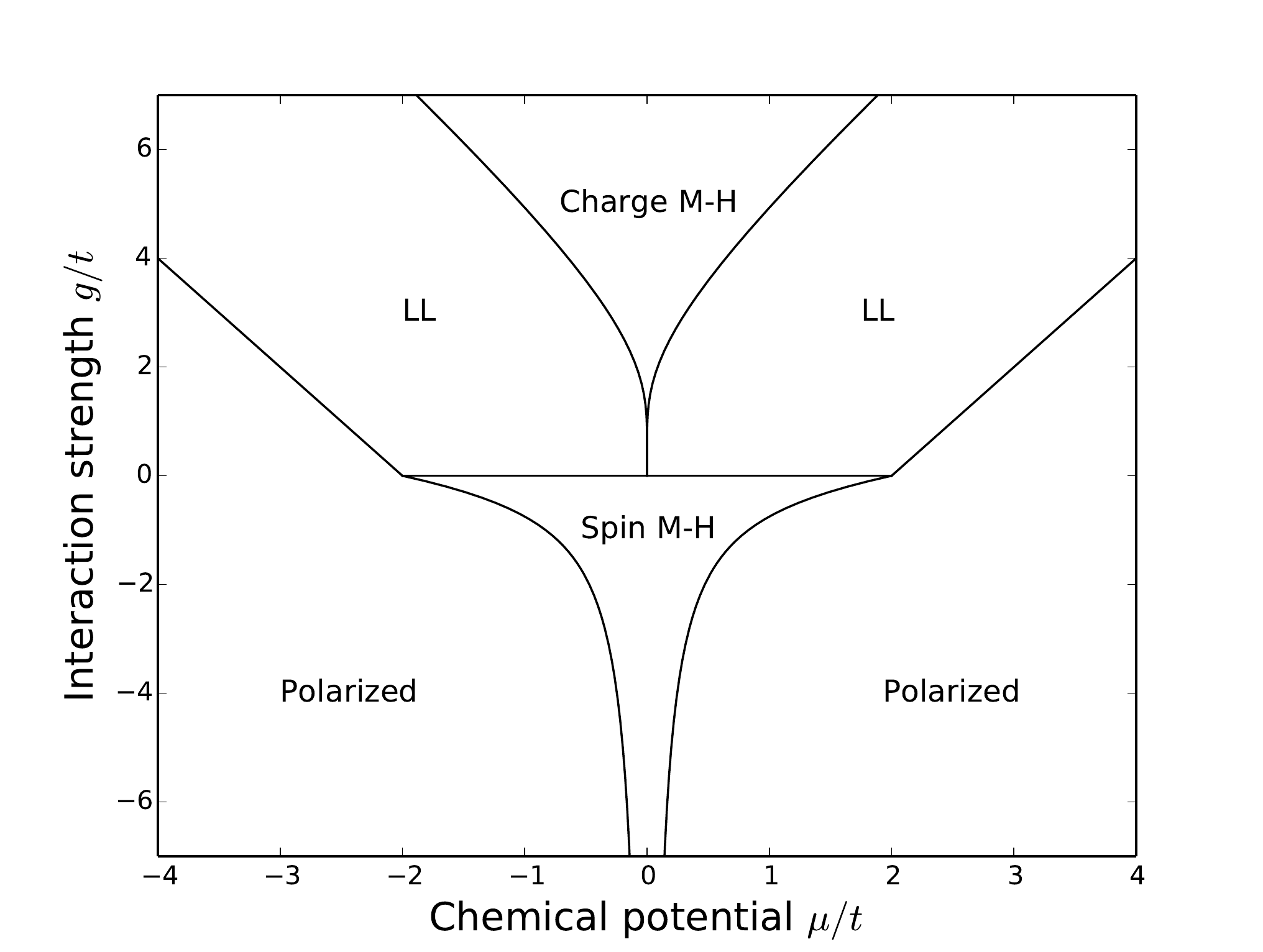}
\end{center}
\caption{Exact phase diagram of the Hubbard model at zero magnetic field, obtained from Bethe Ansatz\protect\citep{essler2005book}. Charge M-H corresponds to the celebrated Mott-Heisenberg phase, opening at arbitrarily low $g>0$. The charge mode is gapped and the electronic density is fixed at half-filling, while the spin mode is free (and its effective model is a $SU(2)$-invariant Heisenberg model). Spin M-H is its equivalent for $g<0$, inversing the role of the two sectors. Both of them are characterized by a central charge $c=1$. Luttinger liquid (LL0 phases) corresponds to two phases of free diluted electrons (with a central charge $c=2$). Polarized phases are trivial phases with totally empty or full wires.}
\label{fig:PDHubbard}
\end{figure}

When $\Delta=0$, our model reduces to the celebrated and well-studied fermionic Hubbard model through the  mapping:
\begin{align*}
c_1 \rightarrow c_\uparrow ~~~~~~
c_2 \rightarrow c_\downarrow.
\end{align*} 
The  $U(1)$ symmetry is restored and the Bethe ansatz method is applicable to solve the model at arbitrary chemical potential. In this Section, we present some known results on this stapled model\protect\citep{essler2005book}.  Its exact phase diagram is displayed in Figure \ref{fig:PDHubbard}.\\
The Hubbard model also has a $SU(2)$ spin symmetry. Introducing $\Delta \ne 0$ breaks this spin rotation symmetry down to the $SO(2) \sim U(1)$ rotations around the $y$-axis:
\begin{equation}
 \begin{pmatrix}
 c_{1} \\ c_{2}
 \end{pmatrix} \rightarrow 
 \begin{pmatrix}
 \cos(\phi) & \sin(\phi) \\
 -\sin(\phi) & \cos(\phi)
 \end{pmatrix}
 \begin{pmatrix}
 c_{1} \\ c_{ 2}
 \end{pmatrix}
 \end{equation} 
which leave the model invariant.
The associated conserved charge is 
\begin{equation}
J_y=i\sum\limits_j \left(c^\dagger_{j,1} c_{j,2}-c^\dagger_{j,2} c_{j,1} \right),
\label{Eq:ConservedCurrent}
\end{equation}
corresponding to the total spin in the $y$ direction.
Non-zero values of $J_y$ indicate a breaking of time-reversal symmetry and the presence of a current between the two wires. A phase with a unique ground state must consequently have $\Braket{J_y}=0$, and the symmetry is actually not broken in any of our phases, in agreement with Mermin-Wagner theorem for continuous symmetries. Nonetheless, as shown in Sec. \ref{subsec-largeg}, the competition between superconductivity and Mott physics in our model will produce orbital currents.

\subsection{Mean-field approach to the $DCI$ phase}\label{sec-meanfield}

While in one dimension, mean-field computations are usually not reliable, they can give us some insights on the physical properties of a model. In the absence of Coulomb interaction $g=0$, the transition between the topological and trivial phase is simply given by $\mu=\pm 2t$ and independent of $\Delta$. Due to the conservation of the fermionic parity in each wire, the only partitioning of the interaction  that one can introduce without explicitly breaking any symmetry is:
\begin{equation*}
g(\rho_1(j)-\frac{1}{2})n_{j,2}+g(\rho_2(j)-\frac{1}{2}) n_{j,1},
\end{equation*}
where $\rho_{1/2}$ is the fermionic density in each wire. In this section, we will focus on ersatz where the density is constant in each wire. Assuming a symmetry between the two wires,  we obtain a simple equation for the transition lines:
\begin{equation}
g_\pm=\frac{\mu \mp 2t}{\rho_\pm-\frac{1}{2}}
\end{equation}
where $\rho_\pm$ is the density at the transition point $\mu=\pm 2t$ for the non-interacting Hamiltonian computed in Eq. \ref{Eq:TransDens}.\\

But one can assume the breaking of the expected symmetry between the two wires $c_1 \leftrightarrow c_2$ and allow for different densities. One has now to compare the potential solutions by minimizing the total energy after solving the following two consistency equations:
\begin{align*}
\rho_{1/2}-\frac{1}{2}=-\int\limits_0^\pi \frac{dk}{4\pi}\frac{-(\mu-g(\rho_{2/1}-\frac{1}{2}))-2t\cos k}{\sqrt{(\mu-g(\rho_{2/1}-\frac{1}{2})+2t\cos k)^2+4\Delta^2\sin^2 k}}.
\end{align*}
There is no simple analytical expression to these solutions but a numerical study reveals the appearance of a set of asymmetrical solutions at finite interaction strength. We find a whole parameter space where there exists an asymmetrical solution whose energy is lower than the symmetrical solution. It is an indicator of the opening of a new phase, and roughly corresponds to the limits of the $DCI$ phase. Nonetheless, as expected, the mean-field argument does not correctly describe its properties, obtained with numerical simulations.   While the mean-field computation predicts a finite difference in densities in each wire even in the thermodynamic limit, numerical simulations assert that the difference in electronic populations between the two wires is only around $2$ fermions, whatever the number of sites we consider. Moreover, while the numerical simulations predict a gapless phase, here the phase is necessarily gapped. However, the mean-field approach has the advantage to simply explain the spontaneous breaking of symmetry between the two wires we observe in numerical simulations: instead of having a single ground state, we obtain a doubly degenerate ground state with fermionic parity (even, odd) and (odd, even). The complete study of the $DCI$ phase and its properties can be found in Section \ref{sec:DCI}.

\section{Bipartite Charge Fluctuations and central charge \label{sec:bipartite}}

The DCI phase that appears at the transition between the polarized and the 4MF phases (which are both gapped) is a gapless phase with a central charge $c=1$. We propose that this phase is an extension of two critical Quantum Ising points. In terms of central charge, there is no difference between a critical $c=1$ bosonic field and two critical $c=\frac{1}{2}$ Majorana fields. Indeed, it is trivial to show that one can divide the bosonic field into two Majoranas or, on the contrary, combine the two Majorana fields to form a bosonic field. Following on the works in Refs \onlinecite{song2011entanglement,rachel2012detecting, song2012bipartite} , we present in this Section a criterion on the bipartite charge fluctuations to differentiate between these two cases, after some reminder on the properties of critical phases in one dimension. Bipartite fluctuations can be measured in ultra-cold atoms through the quantum gas microscope \protect\cite{Bakr2009,Sherson2010}, and in real materials from density-density correlation functions \protect\cite{song2012bipartite} or though capacitance measurements \protect\cite{Petrescu2014}. In spin analogues, the bipartite fluctuations can be measured through spin-spin correlation functions.

\subsection{Central charge and bipartite charge fluctuations}\label{subsec:bipartite-c=1}

In one dimension, conformal invariance of critical models has led to a number of progress, in terms of expressing properties of physical effects in solvable models. One of the fundamental object of a conformally invariant theory is its central charge $c$. It is related, for example, to the scaling of the ground state energy for finite systems or the scaling of the entanglement entropy. We will use the latter to compute this central charge, in order to discriminate between possible critical models, as the entropy is easily computed in DMRG.

We remind the reader the definition of the entanglement entropy for a non-degenerate ground state. Let $\Ket{\phi}$ be the ground state wave function and $\rho=\Ket{\phi}\Bra{\phi} $ the associated density matrix. Let separate our system into two connex parts $A$ and $B$, let $A$ be of length $l$. We define the reduced density matrix $\rho_A$ of $A$ by $\rho_A=\text{Tr}_B (\rho)$, where $\text{Tr}_B$ is the partial trace on $B$. We define the entanglement entropy $S_A$ of $A$ by:
\begin{equation}
S_A=-\text{Tr}(\rho_A \log( \rho_A)).
\end{equation}
For a periodic system , the entropy scales as \protect\cite{CalabreseCardy}:
\begin{equation}
 S_A(l)=\frac{c}{3} \log\left(\frac{L}{\pi}\sin(\frac{l \pi}{L})\right) + {\cal O}(1).
 \end{equation}
 In this expression, $l$ represents the size of the sub-region $A$ while tracing out the region $B$ (the total system has the length $L$). Gapped phases are not conformally invariant and, as a result, have a "central charge" $c=0$, while gapless phases and critical points have usually non trivial central charge. Of relevance in this paper are the central charges of Quantum Ising or a free wire of Majorana fermions, $c=\frac{1}{2}$, and of a LL or a free scalar bosonic mode, $c=1$. When there exists several independent modes, this central charge is additive. Finally, this central charge can be challenging to compute in finite size system, if one has no a priori knowledge of the critical theory and in particular on the mapping between critical theory and original model. Indeed, it can be numerically challenging to extract the exact logarithmic contribution without taking into account the precise finite-size contributions. \\
Another entanglement observable we will be interested in is the scaling of the bipartite charge fluctuations of $A$. Let $Q_A$ be the charge operator of the subsystem $A$, then
\begin{equation}
F_A=\text{Tr}(Q_A^2 \rho_A)-\text{Tr}(Q_A \rho_A)^2=\Braket{Q_A^2}-\Braket{Q_A}^2.
\end{equation}

We present a summary of some results detailed in \onlinecite{song2012bipartite}. Differences of scaling in $F_A$ allow for differentiating between gapped and gapless phases. In particular, in the case of a one-channel free fermion model with $U(1)$ symmetry (a $c=1$ model), the bipartite charge fluctuations scale logarithmically with the size of $A$ and can be computed from the underlining conformal theory. With $K$ the LL parameter of the model, one obtains:
\begin{equation}
F_A(l)=\frac{K}{\pi^2} \log \frac{l}{\alpha} + {\cal O}(1). \label{Eq:FAU(1)}
\end{equation}
$\alpha$ is the short distance cut-off. In a similar one channel model of gapped fermions, we will expect $F_A(l)$ to scale as $al+b + {\cal O}(1)$, where $a$ and $b$ will be non-universal numbers.

A proof can be given \protect\cite{song2011entanglement} : let $\phi$ be a critical real bosonic field. We recall from bosonization (more details will be given in Section IV) that the charge density of a fermionic field in one dimension can be expressed as $-\frac{\partial_x \phi}{\pi}$. One can then rewrite $F_A$ in function of the $\phi$ field:
\begin{equation}
\pi^2 F_A=\iint_{[0,l]^2}dx dy \Braket{\partial_x \phi(x) \partial _x \phi(y)}_{\text{c}} \approx \Braket{(\phi(l)-\phi(0))^2}_{\text{c}}
\end{equation}
To evaluate these correlators, one can either consider directly the OPE of the primary fields $\partial_z \phi$ or the OPE of the vertex operators $e^{i\phi}$, or just the correlator of free bosons in a conformal theory (we note $z=\tau + ix$ and $\omega=\tau'+iy$, with $\tau$ and $\tau'$ imaginary time considered equal here):
\begin{align*}
\partial_z \phi(z) \partial_\omega \phi(\omega) &= -\frac{1}{4\pi \varepsilon}\frac{1}{(z-\omega)^2} \\
\Braket{\phi(x) \phi(0)} &\approx -\frac{1}{2\pi \varepsilon} \log(|x|).
\end{align*}
Getting rid of unphysical terms due to the natural divergences of the theory, we find a term proportional to $\log l$. To identify the coefficient $\varepsilon$, only free parameter of the critical theory, one can look at physical observables such as the compressibility to obtain Equation $\ref{Eq:FAU(1)}$ \protect\cite{song2011entanglement}.\\

This result can be extended to a wire with several channels and a central charge $c=m$, $m\in \mathbb{N}$, corresponding to $m$ gapless bosonic modes and possibly other gapped modes described by Sine-Gordon models. For bipartite charge fluctuations that are quadratic in the bosonic fields, it is shown in Appendix \ref{app:bipartite-c=m} that the logarithmic term in  bipartite charge fluctuations is always positive.

\subsection{Bipartite charge fluctuation for a $c=\frac{1}{2}$ model}

We are now interested in computing the charge fluctuations in the case of a $c=\frac{1}{2}$ model. We refer the reader to Ref.  \onlinecite{CFTSenechal1996} for a review on conformal field theory.

Before entering into the details of the bipartite fluctuations, there is a first point to address. In the Kitaev model (and its counterpart Quantum Ising), the total charge (total spin along $z$) is no longer a good quantum number and a conserved quantity. In particular, the ground state has no longer a proper electron number but is a superposition of states with a different number of particles. This has strong consequences on the bipartite charge fluctuations: there are no longer any symmetry between $F_A(l)$ and $F_A(L-l)$. The quickest way to show this is to consider the two limiting cases $F_A(0)$ and $F_A(L)$. While $F_A(0)$ is still $0$, $F_A(L)$ is no longer equal to $0$ as the charge is no longer fixed on the whole wire. As a consequence, the standard trick from CFT to take into account finite size effects, replacing the length $l$ of the segment we consider by $\frac{L}{\pi} \sin(\frac{l \pi}{L})$ is only valid for the direct conformal contributions. Consequently, we start by considering an infinite wire. Moreover, we can expect a non-zero linear contribution to $F_A$, and not just a logarithmic one. This change breaks the symmetry that existed between the bipartite charge fluctuations and the entanglement entropy for standard Luttinger Liquids. \\

We will prove in this Section that the sign of the logarithmic contribution to the charge fluctuations is {\it negative}. The change in the behavior of these fluctuations comes both from the different underlying critical theory, but also from the difference on how to express the fermionic density in terms of the primary fields. Indeed, the difference will subsist in the case of a $c=2 \times \frac{1}{2}$ theory. We propose two different proofs for the computation of the charge fluctuations. \\

The first one is simply based on exact computation on the regularized lattice model. Details can be found in Appendix \ref{app:bipartite-c=1/2}. One can recover the exact expression of the linear coefficient for all $\Delta/t$, using Fej\'{e}r theorem:
\begin{align}
F_A(l)&=\frac{|\Delta]}{2|\Delta|+2t} l + {\cal O}(\log l). \label{Eq:Fejer}
\end{align}
To obtain the sub-dominant logarithm coefficient, a more involved computation is required. For $\Delta = t$, one can proceed to the complete computation and obtain:
\begin{equation}
F_A(l)= \frac{l}{4}-\frac{1}{2\pi^2}\log(l)-\frac{\gamma_\text{euler}+2\log(2)}{2\pi^2}+{\cal O}(1).
\end{equation}
The logarithmic contribution for the bipartite charge fluctuations are this time negative. Numerical computations of the relevant integrals confirm that the results stand for all $\Delta/t \neq 0$.

We can also recover this result directly from the underlying conformal theory. The critical conformal theory of Quantum Ising can be expressed as a theory of a free real (Majorana) fermions $\psi$, as shown in Section II. We only consider the point $t=\Delta$ for simplicity, but the analysis stands at all $\Delta \neq 0$.

We can reformulate the Hamiltonian in the following way. Let $\gamma_{j,b}=\alpha_{2j}$ and $\gamma_{j,A}=\alpha_{2j+1}$ the two different species of Majorana in the wire.
\begin{equation}
H_K=it\sum\limits_j \gamma_{j,B}(\gamma_{j+1,A}-\gamma_{j,A})-i\frac{\delta \mu}{2}\sum\limits_j\gamma_{j,A} \gamma_{j,B}.
\end{equation}
We first go to continuous limit $\gamma_{j, A/B} \rightarrow \sqrt{2 \alpha}\gamma_{A/B}(x)$, where $\alpha$ is the lattice spacing. Then we introduce two chiral fermions:
\begin{align*}
\gamma_R&=\frac{\gamma_B-\gamma_A}{\sqrt{2}}\\
\gamma_L&=\frac{\gamma_A+\gamma_B}{\sqrt{2}}.
\end{align*} 
Posing $v=2t\alpha$ and $m=\alpha \delta \mu$, the Hamiltonian of the system is now given by:
\begin{equation}
H_K=\int dx  \frac{iv}{2}\left( \gamma_{L}\partial_x \gamma_{L}-\gamma_{R}\partial_x \gamma_{R} \right)-im\gamma_{L} \gamma_{R}. \label{EQ:CFT-Ising}
\end{equation}
At $m=0$, one can identify this Hamiltonian with its conformal action counterpart. Introducing $z=\tau+ix$, $\gamma_L$ and $\gamma_R$ corresponds to the holomorphic and anti-holomorphic part of the conformal field. The action is given by:
\begin{equation}
S=\varepsilon \int dz d\bar{z} \gamma_L \partial_{\bar{z}} \gamma_L + \gamma_R \partial_z \gamma_R.
\end{equation}

$\varepsilon$ is the critical energy scale we will check with correlation functions.

\begin{figure}
\begin{center}
\subfigure[Linear coefficient of the charge fluctuations.]{\includegraphics[width=0.5\textwidth]{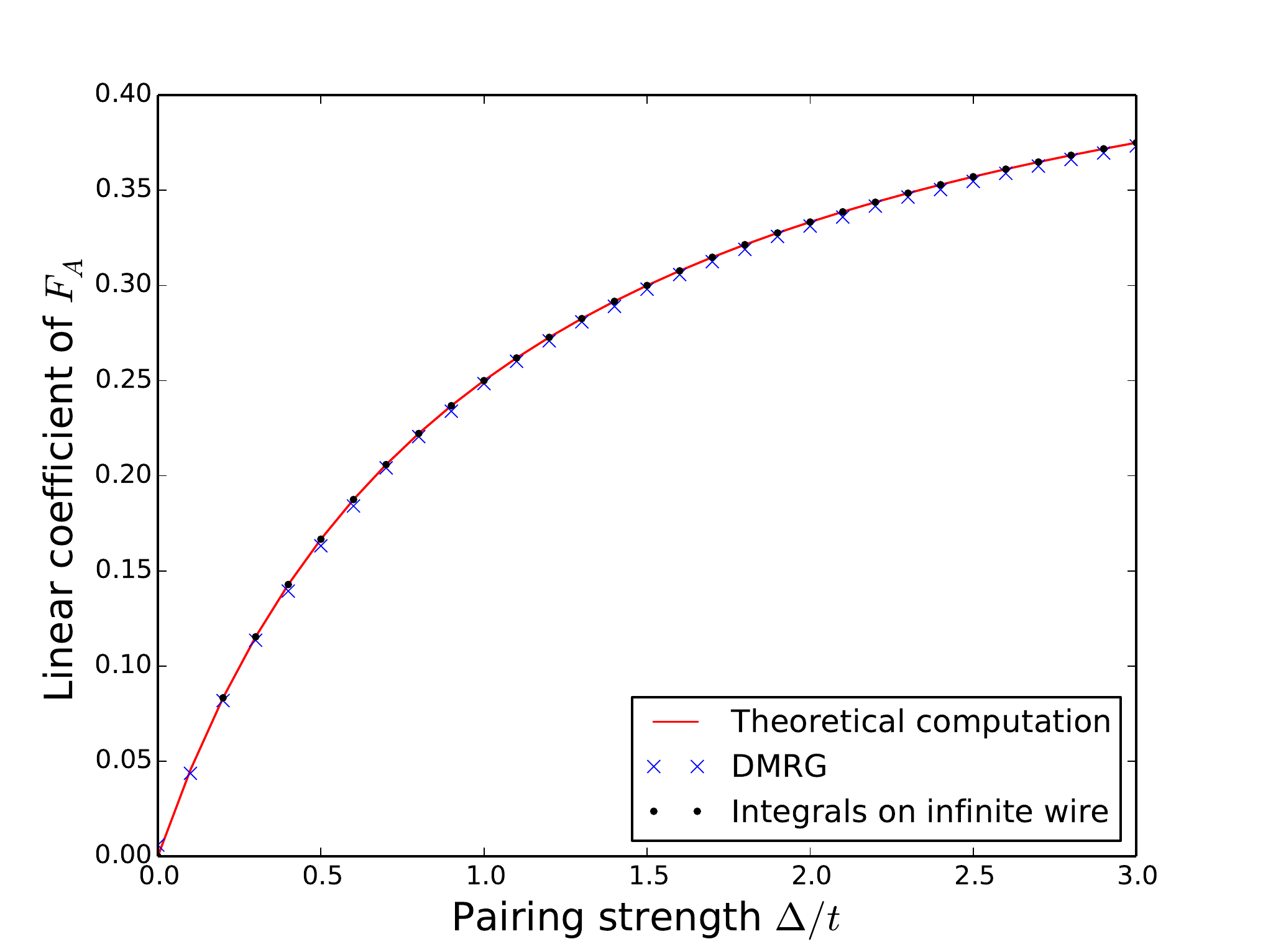}}
\subfigure[Logarithmic coefficient of the charge fluctuations.]{\includegraphics[width=0.5\textwidth]{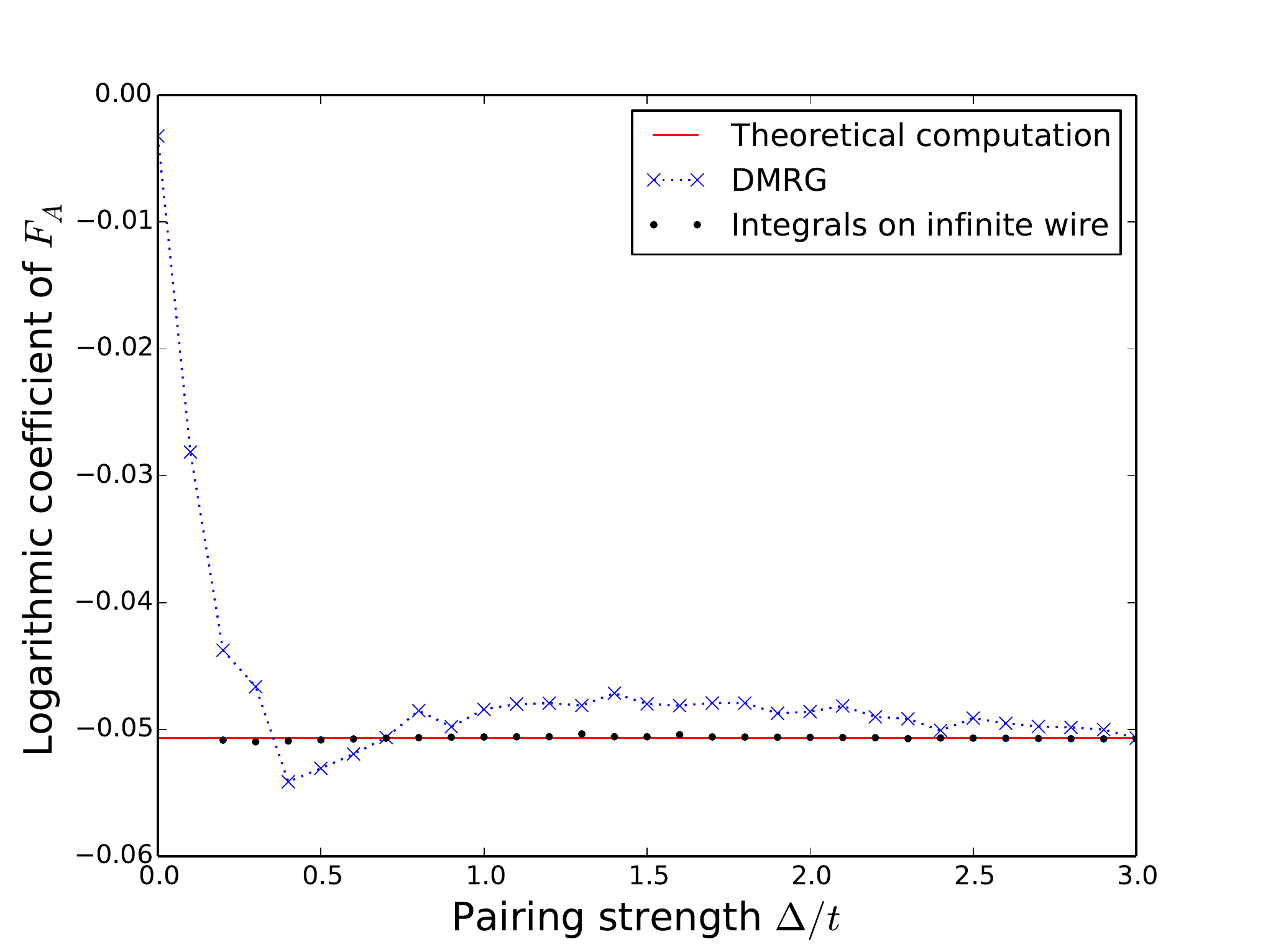}}
\end{center}
\caption{(Color online) Linear and logarithmic contributions to the bipartite charge fluctuations as a function of $\Delta/t$ for a single Kitaev wire at the transition point, from theoretical computations and DMRG simulations \protect\cite{bauer2011, dolfi2014matrix} with a $90$-sites wire. Linear contributions are easily extracted from the simulations and match very well the theoretical values. On the other hand, logarithmic contributions are more complex to extract and convergence is slower. We observe yet a good agreement for $\Delta/t>0.3$. Below this value, finite-size effects cannot be neglected. Fit have been realized using the infinite wire form of the fluctuations, as, at long range, there is a precision loss due to the dominant linear term.}
\label{fig:FAKit}
\end{figure}

The fermionic density operator (i.e the $\sigma^z$ field for the spins) can be written in terms of primary fields as $i\gamma_L(z) \gamma_R(\bar{z})$, up to constant terms that will disappear because we are interested in the connected correlators. One can consequently rewrite:
\begin{equation}
F_A(l)=-\iint\limits_{[0,l]^2} dxdy \Braket{\gamma_L(ix) \gamma_R(-ix)\gamma_L(iy) \gamma_R(-iy)}_c
\end{equation}
Using the Operator Product Expansion (OPE)\protect\cite{CFTSenechal1996, BosonizationGogolin1998}  $\gamma_L(z) \gamma_L(w) = \frac{1}{2 \pi \varepsilon} \frac{1}{z-w}$ for the Majorana field directly yields the result:
\begin{align}
F_A(l)&=-\iint\limits_{[0,l]^2} dxdy \frac{1}{4\pi^2 \varepsilon^2} \frac{1}{|x-y|^2} \nonumber \\
&\approx -\frac{1}{2\pi^2\varepsilon^2} \log l + \alpha l + \beta.
\end{align}
The minus sign in front of the logarithmic contribution can be understood from the $(i)^2 = -1$ pre-factor in Eq. (23) stemming from the definition of the (electron) density operator as $i\gamma_L(z)\gamma_L(w)$.
As in the computation of Section IIIA, we directly got rid of the unphysical divergences of the theory, that are due to the absence of ultraviolet cut-off. $\alpha$ and $\beta$ are a priori non-universal constants that arise from the integration, and are linked to the cut-off of our theory. The results of our two computations coincide for $\varepsilon=1$. This can be confirmed by the computation of the correlation function corresponding to $i\gamma_L(ix) \gamma_L(iy)$ in the original Bogoliubov particles language. One can show that the coefficient of the leading term corresponding to the OPE expansion of the CFT corresponds indeed to $\varepsilon=1$ and, moreover, that it does not depend on $\Delta/t$. In Kitaev's wire, the logarithmic contributions to the bipartite charge fluctuations are actually independent from the ratio $\Delta/t$ as long as $\Delta \neq 0$. Figure \ref{fig:FAKit} provides a comparison between the exact results and some numerical computations using DMRG with Matrix Product States (MPS) from the ALPS project code\protect\cite{bauer2011, dolfi2014matrix}. While the agreement is nearly perfect for the linear contribution, the logarithmic contribution is more complex to catch and slow to converge. In particular, for $\Delta/t \le 0.2$, finite-size effects are far from negligible.

\begin{figure}
\begin{center}
\includegraphics[width=0.5\textwidth]{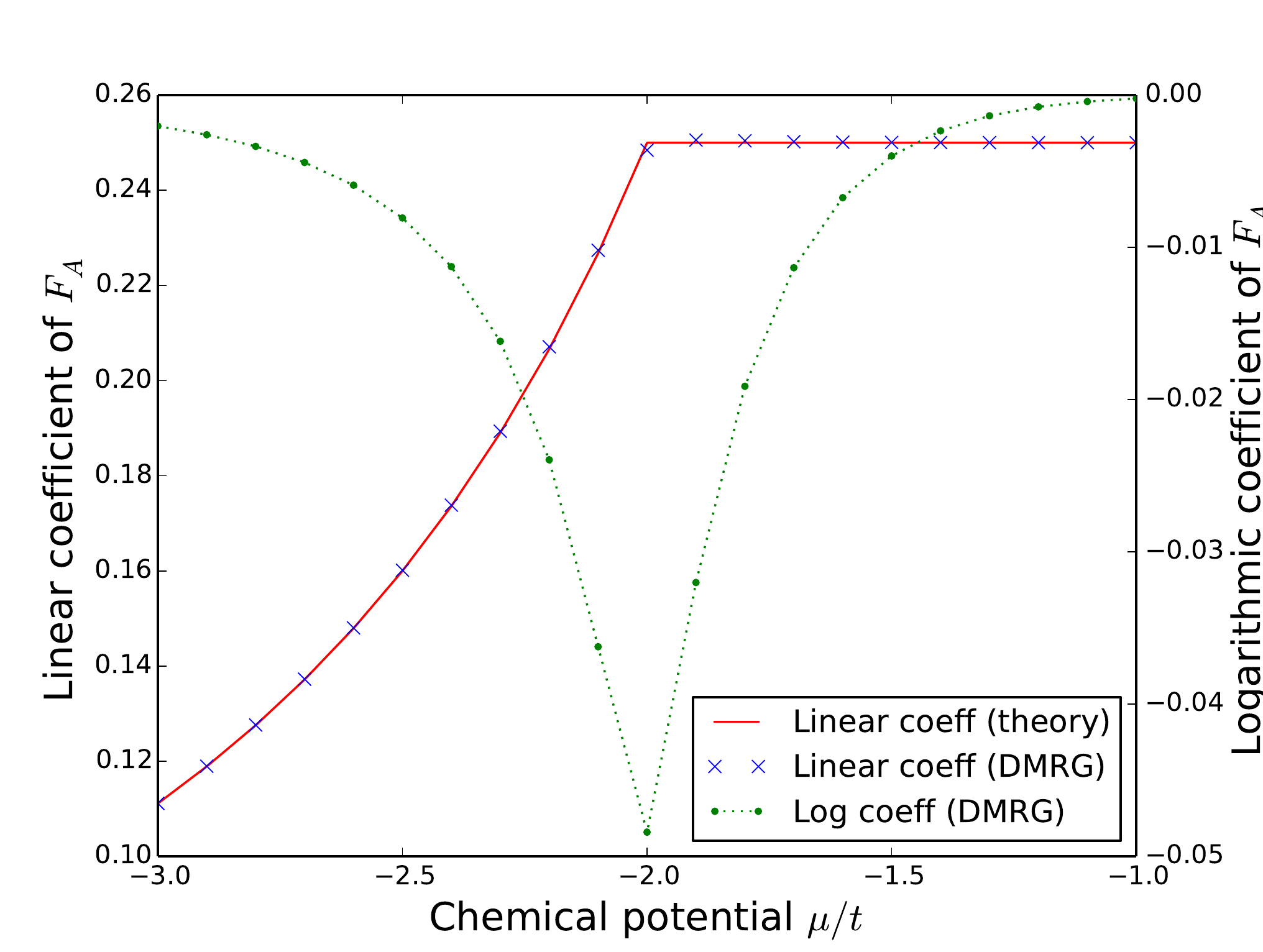}
\end{center}
\caption{(Color online) Linear and logarithmic contributions to the bipartite charge fluctuations as a function of $\mu/t$ for a single Kitaev wire across the transition point, at $\Delta=t$, from theoretical computations and DMRG simulations with a $90$-site wire. Linear contributions are easily extracted from the simulations and match very well the theoretical values. This coefficient becomes constant at $\mu=-2t$. Logarithmic contributions increase as we reach the transition, and are maximum at this point.}
\label{fig:FAMassKit}
\end{figure}

Finally, Figure \ref{fig:FAMassKit} presents the evolution of the two coefficients as we cross the transition line for one Kitaev wire at $\Delta=t$. The introduction of the mass term $m$ in Eq.\ref{EQ:CFT-Ising} formally cuts the logarithm at long range. The logarithmic contribution then goes to zero smoothly away from quantum criticality. The change in the linear coefficient also marks the transition. The exact expression can be found in Appendix \ref{app:bipartite-c=1/2}, obtained from the lattice model using Fej\'{e}r theorem. It is constant for $|\mu|\le 2t$.

\section{Close to half-filling}

In this Section, we are interested in the physics of the complete model of the two wires close to the line $\mu=0$, corresponding to half-filing. We present an analytical approach to the problem, assisted by numerical verifications using both Exact Diagonalization (ED) and Density Matrix Renormalization Group (DMRG). Computations to check convergence of the different parameters have been pushed up to 150 fermions per wire (a total of 300 fermions).

\subsection{Bosonization at half-filling}\label{subsec:boson}

We will proceed with a standard Abelian bosonization scheme \protect\cite{QP1DGiamarchi2004}, considering both $\Delta$ and $g$ as perturbations. The fundamental property behind bosonization is the correspondence between the excitations of a fermionic system above the Fermi surface and the excitations of a well-chosen real bosonic field.  

First we will assume that we are at half-filling and consequently far from the bottom of the energy band in each wire. We can then linearize the spectrum around the Fermi energy and separate each fermion field into its left- and right-moving part:
\begin{equation}
c_{j,\sigma} = e^{ik_F j} c_{1,j,\sigma} + e^{-ik_F j}c_{-1,j,\sigma}.
\end{equation}
Here, $\sigma$ is the index of wire ($\sigma=1, 2$) and $r=1 (-1)$ represents the right-(left-)moving electrons. Going to the continuum limit and linearization of the hopping term gives:
\begin{equation}
-iv_F \sum\limits_\sigma \int dx \left( c^\dagger_{1,\sigma}(x) \partial_x c_{1,\sigma}(x)  - c^\dagger_{-1,\sigma}(x)  \partial_x c_{-1,\sigma}(x)  \right)
\end{equation}
$v_F=2t \alpha \sin(k_F)$ is the Fermi velocity, $\alpha$ is a short distance cut-off we introduce to go from a lattice theory to a continuous theory (Conventionally, $\alpha$ corresponds to the distance between two sites of the lattice) and $k_F=\pi/2$ the Fermi momentum. We then introduce two real scalar bosonic fields $\phi_\sigma$ and their conjugate field $\theta_\sigma$ corresponding to the excitations of each wire, and four Majorana fermions $U_{r, \sigma}$ that will serve as Klein factors. They verify the following commutation rules:
\begin{equation}
[\phi(x), \theta(x')]=i\frac{\pi}{2}\text{sgn}(x'-x)~~\text{and}~~\{U_r, U_{r'}^\dagger\}=2\delta_{r,r'}.
\end{equation}
The well-known mapping between the fermions and the bosonic fields is given, in the thermodynamic limit, by:
\begin{equation}
c_{r,j, \sigma}=\frac{U_{r, \sigma}}{\sqrt{2 \pi \alpha}} e^{-i(r\phi_{j,\sigma}-\theta_{j, \sigma})}.
\end{equation}
After some computations, using the convention $U^\dagger_L U^\dagger_R=-i$, the Hamiltonian  at $g=0$ is:
\begin{equation}
H = \sum\limits_\sigma \int dx \frac{v_{F}}{2\pi}((\partial_x \theta_\sigma)^2 + (\partial_x \phi_\sigma)^2)+\frac{2\Delta \sin(k_F)}{\pi\alpha}\cos(2\theta_\sigma)
\end{equation}
with $v_F=2ta\sin(k_F)$. We then introduce the charge $\phi_+ =\frac{\phi_1+\phi_2}{\sqrt{2}}$ and spin ( relative charge mode in our case) modes $\phi_- =\frac{\phi_1-\phi_2}{\sqrt{2}}$. The density-density interaction term renormalizes the Fermi velocities and the Luttinger parameters of both bands, and induces two cosine terms. The final Hamiltonian, including terms generated in the Renormalization Group (RG) process, is given by:
\begin{multline}
H=\sum\limits_{\varepsilon=\pm}\int dx \frac{v_{F,\varepsilon}}{2\pi}(K_\varepsilon(\partial_x \theta_\varepsilon)^2 + K_\varepsilon^{-1}(\partial_x \phi_\varepsilon)^2)\\
+ \frac{g_\varepsilon}{\alpha^2} \cos(2\sqrt{2} \phi_\varepsilon)-\frac{\Delta^{(2)}_\varepsilon}{\alpha^2} \cos(2\sqrt{2} \theta_\varepsilon)\\
+ \frac{\Delta^{(1)}}{\alpha^2}\cos(\sqrt{2}\theta_+ ) \cos(\sqrt{2}\theta_-).
\label{Eq:BosonHalf}
\end{multline}
Scaling dimensions and bare values of the different parameters can be found in Table \ref{tab:DimensionBasicModel}. At this order of RG theory, we do not consider the renormalization of the Fermi velocities. $K_+$ and $K_-$ are the Luttinger parameters. $g_+$ and $g_-$ appear due to the Coulomb coupling between the two wires. Both are present in Hubbard model and are responsible for the Mott-Heisenberg phases. The pairing $\Delta$ plays now the role of a coupling $\Delta^{(1)}$ between the two charge and spin sectors, that cannot be a priori separated. $\Delta^{(2)}_+$ and $\Delta^{(2)}_-$ are not initially present in the bare Hamiltonian, but are generated under RG by $\Delta^{(1)}$ . In a diagrammatic language, they correspond to second order contributions in $\Delta$.
\begin{table}
\begin{center}
\begin{tabular}{|c|c|c|}
\hline
 Term & Dimension &Bare value  \\ \hline
 $v_{F, \pm}$ & - & $v_F \sqrt{1\pm \frac{g}{\pi v_F}}$ \\ \hline
 $K_\pm$ & - &$ K_\pm(0)=\frac{1}{\sqrt{1\pm \frac{g}{\pi v_F}}}$ \\ \hline
$\cos(\sqrt{2}\theta_-) \cos(\sqrt{2}\theta_+)$ & $\frac{1}{2}(K_+^{-1}+K_-^{-1})$ & $\Delta^{(1)}(0)=\frac{4 \Delta \alpha}{\pi }$   \\ \hline
$\cos(2\sqrt{2}\theta_+)$ & $2K_+^{-1}$ & $\Delta^{(2)}_+(0)=0$   \\ \hline
$\cos(2\sqrt{2}\theta_-)$ & $2K_-^{-1}$ & $\Delta^{(2)}_-(0)=0$  \\ \hline
$\cos(2\sqrt{2}\phi_+)$ & $2K_+$ & $g_+(0)=\frac{-g}{2\pi^2}$   \\ \hline
$\cos(2\sqrt{2}\phi_-)$ & $2K_-$ & $g_-(0)=\frac{g}{2\pi^2}$  \\ \hline
\end{tabular}
\end{center}
\caption{Dimensions of the different terms of the bosonized model, and bare values in the RG flow.}
\label{tab:DimensionBasicModel}
\end{table}

We define the renormalization length as: $\alpha(l)= \alpha e^l$. Bosonization consequently allows us to recover the following renormalization flow equations, including all relevant orders:
\begin{align}
\frac{dK_\pm}{dl}&=-\frac{2 \pi^2 g_\pm^2}{v_{F,\pm}^2} K^2_\pm + \frac{2 \pi^2(\Delta^{(2)}_\pm)^2}{v_{F,\pm}^2}+\frac{\pi^2 (\Delta^{(1)})^2}{4 v_{F,+} v_{F,-}} \nonumber \\
\frac{dg_\pm}{dl}&= (2-2K_\pm)g_\pm \nonumber \\
\frac{d \Delta^{(1)}}{dl}&=(2-\frac{1}{2}(K_+^{-1}+K_-^{-1}))\Delta^{(1)} \nonumber \\
\frac{d \Delta^{(2)}_\pm}{dl}&=(2-2K_\pm^{-1})\Delta^{(2)}_\pm+\frac{2\pi^2(\Delta^{(1)})^2}{v_{F,\mp}}.
\end{align}

Dimensional analysis of these equations discriminates three different phases at half-filling, to be analyzed below. To qualitatively compare the effects of $g_\varepsilon$ and $\Delta^{(1)}$, we compare the bare value of the latter, an a priori strongly relevant coupling, and the effective mass $m_g$ obtained by Bethe-Ansatz in the Hubbard model (in other words the Mott gap in the charge sector), at low coupling~\protect\cite{essler2005book}
\begin{align*}
\frac{m_g}{t} &= -2 + \frac{|g|}{2t}+2\int\limits_0^{\infty}d\omega \frac{J_1(\omega)e^{-\frac{|g| \omega}{4t}}}{\omega} \\
&\approx \frac{4}{\pi} \sqrt{\frac{|g|}{t}} e^{-\frac{2 \pi t}{|g|}} ~~~~~~\text{for $|g|\ll t$.}
\end{align*}
$J_1$ is the Bessel function of the first kind.
\begin{itemize}
\item If $\Delta \gtrsim |m_g|$,  $\Delta^{(1)}$ dominates the $g_{\pm} \cos(2\sqrt{2}\phi_{\pm})$ terms and goes to strong coupling. Both $\theta_{\pm}$ modes become massive and are locked to the minima of the $\Delta^{(1)}$ term. By continuity with the topological phase at $g=0$, we expect this strong coupling fixed point to correspond to the SPT phase presenting four Majoranas, or $4MF$ phase.
\item If $|m_g| \gtrsim \Delta$ and $g>0$, $g_+$ is renormalized to large coupling before $\Delta^{(1)}$ reaches significant values. $g_-$ is irrelevant and renormalized to $0$. $\phi_+$ is consequently locked to $0~ [\pi / \sqrt{2}]$, corresponding to a Mott ordered phase, and $\Delta^{(1)}$ vanishes at this fixed point.  $\Delta^{(2)}_-$ is still relevant and acquires a non-zero value in the initial steps of the renormalization. It consequently gaps the spin sector and both modes $(\phi_+, \theta_-)$ are eventually locked. This fixed point describes the $MI-AF$ phase.
\item If $|m_g| \gtrsim \Delta$ and $g<0$, the reasoning is the same as for $g>0$ but for an inversion of the charge and the spin sector.  $(\phi_-, \theta_+)$ are locked, describing the $MI-F$ phase.
\end{itemize} 

As long as we stay close to half-filling, one can use the same bosonization scheme to determine the effects of a chemical potential. Indeed, one only needs to add a term:
\begin{equation*}
- \frac{\mu \sqrt{2}}{\pi} \partial_x \phi_+,
\end{equation*}
in the Hamiltonian. The effect of this term is two-fold: it reduces the effective dimension of $\cos(2\sqrt{2} \phi_+)$ and renormalizes the Fermi velocity~\protect\cite{Schulz1980}. When the renormalized Fermi speed approaches $0$, it indicates that we are too far from half-filling and that the spectrum is no longer linear, leading to the breakdown of the bosonization approximation. We will summarize the effect of the chemical potential on each of the previously obtained phases:
\begin{itemize}
\item If $\Delta \gtrsim |m_g|$, none of the fixed operators includes a $\phi_+$ term, meaning that no transition occurs before we reach the bottom of the band and the bosonization procedure breaks down.
\item If $|m_g|\gtrsim \Delta$ and $g>0$, the umklaap term controlled by $g_+$ starts oscillating. In the Hubbard model  occurring at $\Delta=0$, $g_+$ is renormalized to zero at a finite ratio  $\frac{\mu}{g}$, corresponding to a vanishing charge gap and a commensurate-incommensurate transition to a gapless Luttinger phase~\protect\cite{Schulz1980}. At finite but small $\Delta$, $\mu$ weakens $g_+$ by reducing its dimension until $\Delta^{(1)}$ dominates the RG process and flows to strong coupling. This leads to a resurgence of the 4 Majorana phase at  finite $\mu$.
\item If $|m_g|\gtrsim \Delta$ and $g<0$, as with the first phase, a transition does not occur in the bosonization range.
\end{itemize} 

We describe in the next two parts the properties of these three phases.

\subsection{Characterization of the 4 Majorana phase}

Based on adiabaticity, we expect the topological properties of the $4MF$ phase to be well-described by the $g=0$, $|\mu| < 2 t$ case, {\it i.e.} two uncoupled Kitaev wires in their topologically non-trivial phases. Hence, four zero-energy Majorana end states should be present and remain uncoupled, corresponding to a fourfold degenerate ground state. We present in this section a few analytical and numerical arguments that support this claim. Following the $\mathbb{Z}_8$ classification by Kitaev and Fidkowski\protect\cite{Fidkowski2010}, the phase is, in this case, simply a symmetry protected topological phase (SPT). Indeed, these Majoranas are not protected against any local interaction: an arbitrarily low pairing term between the two wires $i|\Delta_\perp|(c^\dagger_{j,1}c^\dagger_{j,2} -c_{j,2} c_{j,1})$ would directly couple the free Majoranas, destroying the phase. 

A first approach consists in considering the perturbative effect of $g$ on the extremity of two Kitaev wires in the topological phase. We assume $t=\Delta$ to get a simpler picture. Using the notation of Section~\ref{sec-kitaev}, we recall that the free Majorana fermions are $\gamma_{1, \sigma}^B$ and $\gamma_{L, \sigma}^A$ (the additional $\sigma$ is the wire index). The interaction term Eq.~\eqref{eq:inter} can be rewritten as: $-\frac{g}{4} \sum\limits_j \gamma_{j,1}^A\gamma_{j,1}^B \gamma_{j,2}^A\gamma_{j,2}^B$. Only terms at least of order $(\frac{g}{t})^L$ will directly couple the free Majoranas. In the thermodynamic limit, we consequently expect the survival of the 4 Majoranas phase.\\

\begin{figure}
\begin{center}
\includegraphics[width=0.5\textwidth]{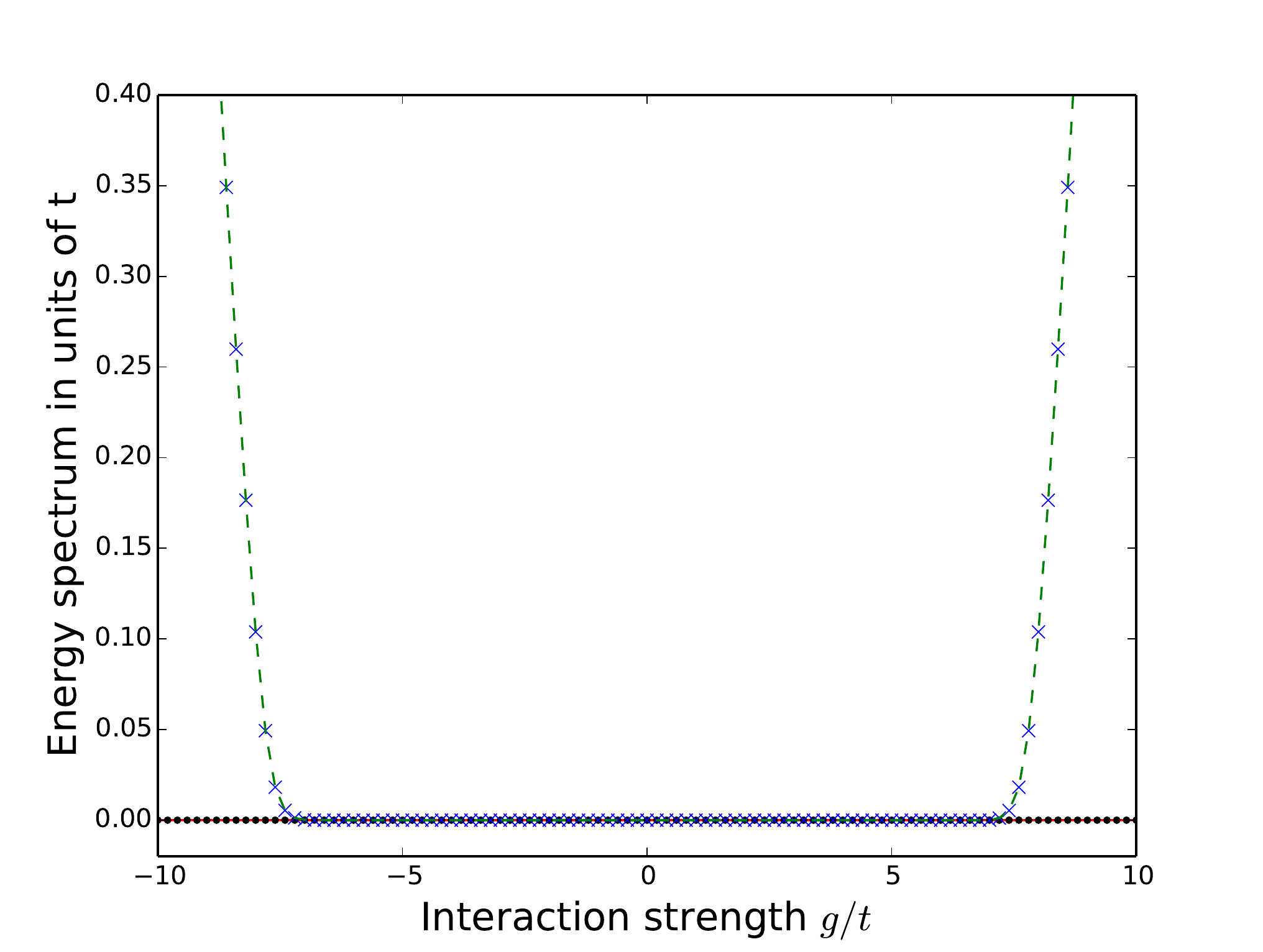}
\end{center}
\caption{(Color online) The first four levels of the energy spectrum on the line $\mu=0$ for $t=\Delta$ and with open boundaries for a 60-sites ladder. We observe a four-fold degeneracy in the topological phase, and a two-fold degeneracy in the Ising phases.}
\label{fig:SpecMu0}
\end{figure}

From a numerical point of view, we have studied several markers for the topological phase. The first is obviously the change in degeneracy going OBC to PBC. While in the latter the ground state present no degeneracies, it is four-times degenerate in the former. The parity of the ground states follows the same rules as in the non interacting case: we go from an odd-odd ground state ($P_1=P_2=-1$) with PBC to a ground state in each parity sector ($P_1=\pm 1$, $P_2=\pm1$) with OBC.  As fermionic parity in each wire is a good quantum number, it is quite easy to observe this degeneracy with both our ED or DMRG simulations. Typical behavior for the energy of the first few levels on the line $\mu=0$ is presented in Figure \ref{fig:SpecMu0}.

\begin{figure}
\begin{center}
\includegraphics[width=0.5\textwidth]{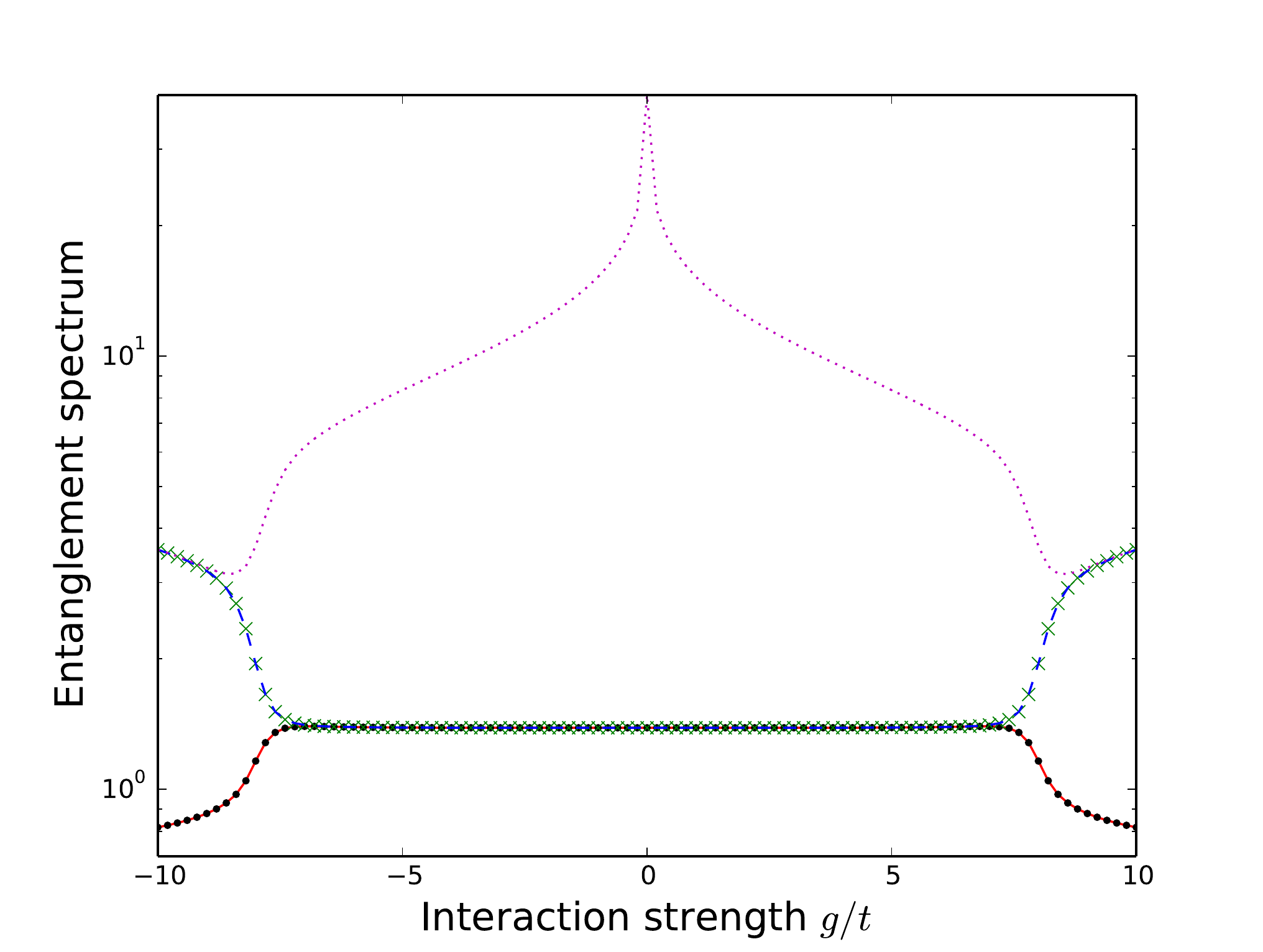}
\end{center}
\caption{(Color online) The first five levels of the entanglement spectrum on the line $\mu=0$ for $t=\Delta$ and periodic boundary conditions for a 60-sites ladder. We observe a four-fold degeneracy in the topological phase, and a two-fold degeneracy in the Ising phase due to the degeneracy of the ground state with PBC.}
\label{fig:EntSpecMu0}
\end{figure}

A second good marker for topology is the degeneracy of the entanglement spectrum \protect\cite{LiHaldane} in the periodic ladder.These eigen-energies are obtained by re-interpreting the eigenvalues of the reduced density matrix 
$\rho_A=\text{Tr}_A(\Ket{\Psi}\Bra{\Psi})$ as Gibbs exponential forms. Here $A$ is simply the half of the wire. A comprehensive and clear point of view of the properties of the entanglement spectrum for topological fermionic phases can be found in Ref. \onlinecite{Turner2011}. The presence of Majorana boundary states translates into a four-fold degeneracy in the entanglement spectrum, as cutting the system is analogous to creating new boundaries. As in the previous numerical probe, this degeneracy can be observed in both ED or MPS simulations. Typical behavior for the entanglement spectrum on the line $\mu=0$ is presented in Fig. \ref{fig:EntSpecMu0}.\\

\begin{figure}
\begin{center}
\includegraphics[width=0.5\textwidth]{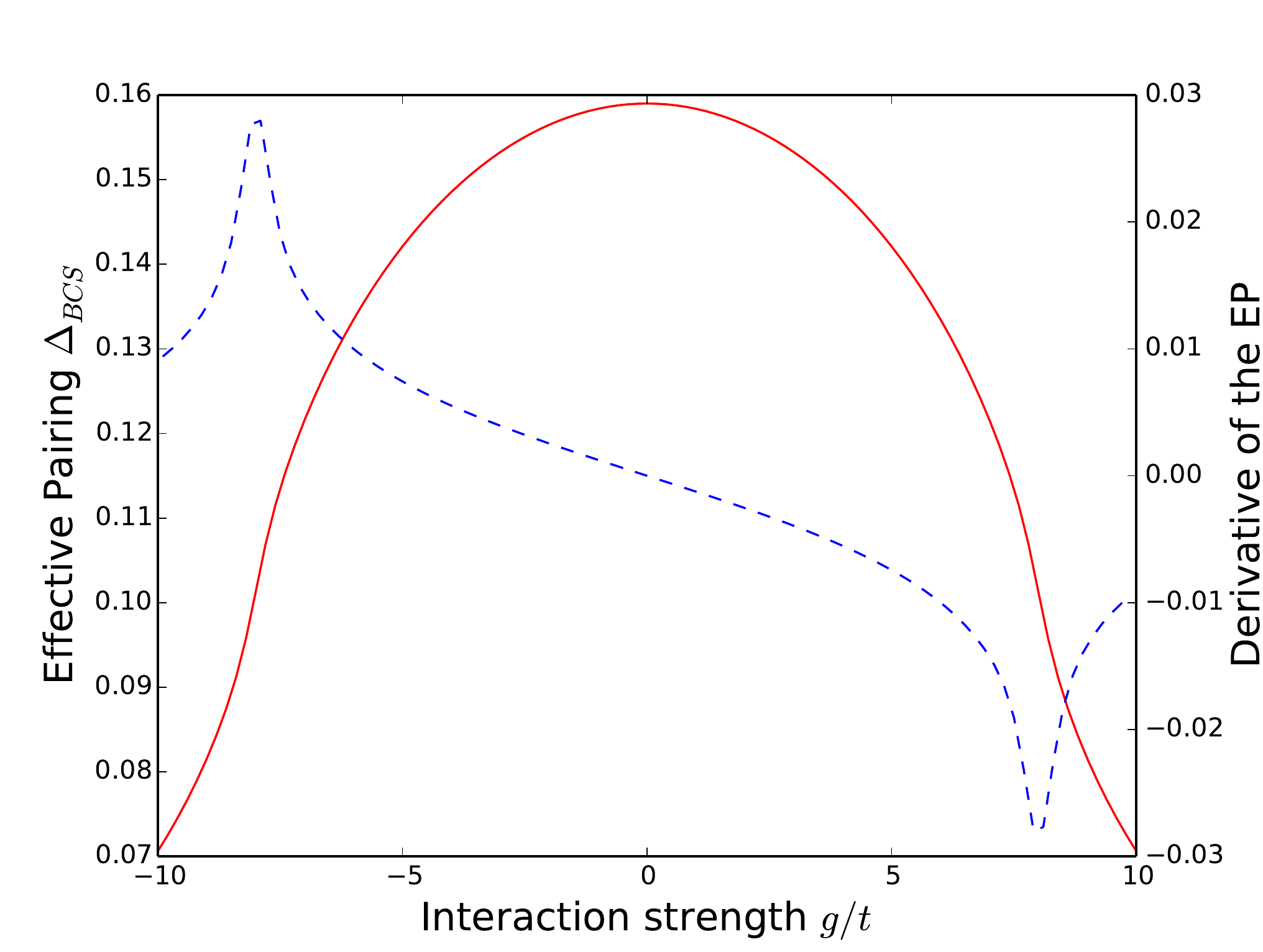}
\end{center}
\caption{(Color online) Effective pairing (in red) and its derivative (in blue) as a function of the interaction strength on the line $\mu=0$ for $\Delta=t$ and periodic boundary conditions for a 60-site ladder.}
\label{fig:EffPairMu0}
\end{figure}

Finally, one can observe a non-local order parameter for Kitaev wire, the effective pairing $\Delta_{BCS}$: 
\begin{equation}
\Delta_{\text{BCS}}=\frac{1}{iL}\sum\limits_{k=0}^\pi \Braket{c^\dagger_k c^\dagger_{-k}}.
\end{equation}
Details of its derivation can be found in Appendix \ref{appsub:Observables}. At $g=0$, it is constant in the topological phase and decreases in the trivial phase, as shown in Figure \ref{fig:EffPaiNoInt}. For numerical simulations, due to finite size effects, there can be no constant term in the topological phase. Nonetheless, one can observe a significant change of behavior in the derivative even at finite size as shown in Figure \ref{fig:EffPairMu0}. The lack of precision reduces the effectiveness of the effective pairing for studying the $DCI$ phase

\subsection{Behavior of the large $g$ phases}

The two $MI$ phases at large $g$ are very similar in behavior. For $\Delta^{(2)}_{\pm} = \Delta^{(1)} = 0$, the effective low-energy Hamiltonian is exactly the same as for the Hubbard model, with a gapless spin sector and a charge Mott gap for $g>0$ (and the opposite for $g<0$). $\Delta^{(2)}_{-}$ ($\Delta^{(2)}_{+}$) however opens a gap in the spin (charge) sector for $g>0$ ($g<0$) and the resulting Mott phases are fully gapped. To obtain the general physical properties of these phases, it is enough to consider a Schrieffer-Wolff transformation into the low-energy subspace at $g\rightarrow \pm\infty$.\\

Let us start with $MI$-$AF$ and $g>0$. Similar to the Hubbard model, we can define spin operators in the natural way: 
\begin{align*}
\sigma_j^z&=c^\dagger_{j,1} c_{j,1}-c^\dagger_{j,2} c_{j,2}\\ \sigma_j^x&=c^\dagger_{j,1}c_{j,2}+c^\dagger_{j,2}c_{j,1}\\
\sigma_j^y&=i(c^\dagger_{j,2}c_{j,1}-c^\dagger_{j,1}c_{j,2}).
\end{align*}
 The effective low-energy Hamiltonian is given by:
\begin{multline}\label{eq:eff}
H_{\text{eff}, g+} =  \frac{t^2-\Delta^2}{2g} \sum\limits_j \sigma_j^z \sigma^z_{j+1} \\+ \frac{t^2 + \Delta^2}{2g}\sum\limits_j \sigma_j^y \sigma_{j+1}^y + \frac{t^2 - \Delta^2}{2g}\sum\limits_j \sigma_j^x \sigma_{j+1}^x.
\end{multline}

Up to a spin-axis rotation, this effective model is nothing but the standard XXZ model. As mentioned in Section IIB, $\Delta$ breaks the $SU(2)$ rotation symmetry of spins of the Hubbard model, preserving only a $U(1)$ rotation invariance around the $y$-axis that can be directly seen in the Hamiltonian. Moreover, as long as $\Delta \neq 0$, we stay in the Antiferromagnetic Ising phase of this model (the N\'{e}el phase, where the anisotropy dominates). Our effective model is consequently gapped in both sectors, and presents a double degeneracy if we have OBC or PBC with an even number of sites. The fermionic density is also fixed at half-filling as long as we stay in this phase. Just as for the Hubbard model, small variations of the chemical potential do not affect the low-energy Hamiltonian, and hence the different observables are left essentially unaffected. A clear physical picture is obtained when $\Delta =t$. The effective model is then a pure Ising model, with trivial ground states:
$\bigotimes_j \Ket{(-1)^j}_j^y$, where $\sigma_j^y\Ket{\pm 1}_j^y=\pm \Ket{\pm 1}_j^y$. This peculiar order in the y-direction is characteristic of the formation of orbital (alternating in this case) currents in the ladder.   Each of these ground state spontaneously breaks the time reversal symmetry, as the transverse operator current
\begin{equation}
J_\perp^N = i \sum\limits_j (-1)^j (c^\dagger_{j,1}c_{j,2}-c^\dagger_{j,2}c_{j,1})
\end{equation} 
acquires a non-zero expectation value.
This spontaneous symmetry breaking is not in contradiction with Mermin-Wagner theorem, as time-reversal is a discrete symmetry ($\mathbb{Z}_2$) in this problem. We want nonetheless to underline that these orbital currents appear in the absence of an explicit flux, and that they are quite unusual as they correspond to coordinated exchange of 4 fermions between the two wires. Ref. \onlinecite{Carr2006} found a similar Orbital Antiferromagnetic phase (OAF), induced by the nearest neighbor interaction $V n_j n_{j+1}$. It is not surprising as $\cos(2\sqrt{2} \theta_-)$, generated by $\Delta$ in the RG process, is also a contribution of $V$. The main difference is in the nature of the spontaneous currents,  as direct hopping between the two wires is allowed in Ref.~\onlinecite{Carr2006}. The occurrence of orbital currents in two dimensions has also attracted some attention in the context of high-Tc superconductors due to the interplay between the magnetism close to the Mott state and the superconductivity\protect\cite{Hsu1991,Chakravarty2001,Varma2000,Kaminski2002}. They can also be induced through the creation of magnetic fields and orbital effects as in superconductors and through artificial gauge fields \protect\cite{Dalibard2011} in ultra-cold atoms \protect\cite{Bloch2008}. For recent examples in two-leg ladder systems, see for example 
Refs. \onlinecite{Muenich,Dhar,AlexKaryn,AlexKaryn2,Marie,Antoine,Sela}. \\
$\mu$ does not change perturbatively the effective Hamiltonian Eq.~\eqref{eq:eff}, as the corresponding term is constant when projected onto the low energy subspace. However, larger $\mu$ of order $g$ are responsible for a resurgence of the $4MF$ phase as discussed above.\\

The case $g<0$ is very similar in its mathematical structure. We define this time anomalous spin operators:
\begin{align*}
\mathfrak{s}_j^z&=c^\dagger_{j,1} c_{j,1}+c^\dagger_{j,2} c_{j,2}-1\\ \mathfrak{s}_j^x&=c^\dagger_{j,1}c^\dagger_{j,2}+c_{j,2}c_{j,1}\\
\mathfrak{s}_j^y&=i(c_{j,2}c_{j,1}-c^\dagger_{j,1}c^\dagger_{j,2}).
\end{align*}  
The corresponding effective Hamiltonian is very similar to the one previously obtained:
\begin{multline}
H_{\text{eff}, g-} = \frac{t^2-\Delta^2}{2|g|} \sum\limits_j \mathfrak{s}_j^z \mathfrak{s}^z_{j+1} - \frac{t^2 + \Delta^2}{2|g|}\sum\limits_j \mathfrak{s}_j^y \mathfrak{s}_{j+1}^y \\
+ \frac{ \Delta^2-t^2 }{2|g|}\sum\limits_j \mathfrak{s}_j^x \mathfrak{s}_{j+1}^x -\mu \sum\limits_j \mathfrak{s}_j^z.
\end{multline}
The physics is the same as for $g>0$, as we can map one to the other with the transformation: $\mathfrak{s}^z_j\rightarrow \sigma^z_j$, $\mathfrak{s}^x_j\rightarrow (-1)^j \sigma^x_j$ and $\mathfrak{s}^y_j\rightarrow (-1)^j \sigma^y_j$
In term of these anomalous spins, we obtain a gapped Ferromagnetic Ising phase at $\mu=0$. Its susceptibility to relative chemical potential (a magnetic field in the language of Hubbard model) is zero. With the chemical potential, the effective model is simply Quantum Ising where it plays the role of the transverse field. \\
Time reversal symmetry is again spontaneously broken in the Ising phase, leading to global currents from one wire to the other. The relevant operator $J_\perp^A$ is obtained by considering the case $\Delta=t$:
\begin{equation}
J_\perp^A = i \sum\limits_j (c_{j,1}c_{j,2}-c^\dagger_{j,2}c^\dagger_{j,1}).
\end{equation}
Ref. \onlinecite{Carr2006} does not observe a similar phase, as $V n_j n_{j+1}$  cannot give a contribution similar to $\Delta^{(2)}_+$. \\

\subsection{Nature of the transitions}

Finally, we provide numerical and analytical arguments for the nature of the different phase transitions. We start by considering the transition between the $MI$-$AF$ and $4MF$  phase. \\

We focus on: $g>0$ and $m_g$ and $\Delta$ are of the same order. First, we argue that the mode $(\phi_-, \theta_-)$ is not affected by the transition. Indeed, $\theta_-$ is still locked to $\theta_- = 0~[\pi /\sqrt{2}]$ and $\Delta^{(2)}_-$ stays relevant at this transition, whether $\Delta^{(1)}$ goes to strong coupling or not. Consequently, the spin sector knows no phase transition. The description of the transition of the charge sector at finite value of $\Delta$ is actually a more challenging problem. Let us start by considering the Hubbard model and the commensurate-incommensurate transition between the Mott phase and the liquid phase. We know from exact solutions that, close to the transition, the universal value of the Luttinger parameter for the charge mode is $K_+=\frac{1}{2}$\protect\cite{Schulz1990, Ren1993}. Branching an arbitrarily low $\Delta$ should not change this picture. 

As the spin mode is supposed to stay gapped, we perform a mean-field approximation in order to separate the two modes:
\begin{align*}
\Delta^{(1)} \cos(\sqrt{2} \theta_+) \cos(\sqrt{2} \theta_-) &\rightarrow \Delta^{(1)} \cos(\sqrt{2} \theta_+)\Braket{ \cos(\sqrt{2} \theta_-)} \\
& \rightarrow \pm \Delta^{(1)} \cos(\sqrt{2} \theta_+).
\end{align*}
The sign depends on the ground state of the spin mode, and has no consequences on the picture of the transition. We will consequently consider it positive. We then proceed to a rescaling $\phi_+ \rightarrow \phi_+/\sqrt{2}$ to reach the refermionizable point. The effective model close to the Hubbard transition line between the Mott-Heisenberg phase and the LL is:
\begin{multline}
H=\int dx \frac{v_{F,+}}{2\pi}((\partial_x \theta_+)^2 + (\partial_x \phi_+)^2)\\
+ \frac{g_+}{\alpha^2} \cos(2 \phi_+)+\frac{\Delta^{(1)}}{\alpha^2} \cos(2 \theta_+),
\end{multline}
where $g_+$ is a small effective interacting term, the effective mass. It corresponds to a $g$ term that has been eventually renormalized by the chemical potential. One can then refermionize the Hamiltonian as done for example in Refs. \onlinecite{Schulz1980,Karyn1999,Lecheminant2002}. To that end, we introduce two chiral fermions:
\begin{equation}
 \psi_{R/L}=\frac{U_{R/L}}{\sqrt{2\pi \alpha}}e^{\mp i(\phi_+ \pm \theta_+)},
 \end{equation} 
where $U_{R/L}$ are Klein factors and $\alpha$ the short distance cut-off of the theory. We place ourselves in the representation where $U^\dagger_R U_L=-i$. We then use the following identification:
\begin{align*}
\cos(2\phi_+)&=i\pi \alpha (\psi^\dagger_R \psi_L-\psi^\dagger_L \psi_R) \\
\cos(2 \theta_+)&= i\pi \alpha (\psi_R \psi_L-\psi^\dagger_L \psi^\dagger_R).
\end{align*} 
The effective Hamiltonian is then given by:
\begin{multline}
H=\int dx(-i v_{F,+})(\psi^\dagger_R \partial_x \psi_R-\psi^\dagger_L \partial_x \psi_L) \\
+\frac{i g_+ \pi}{\alpha} \int dx (\psi^\dagger_R \psi_L-\psi^\dagger_L \psi_R) + \frac{i \Delta^{(1)} \pi}{\alpha} \int dx (\psi_R \psi_L-\psi^\dagger_L \psi^\dagger_R).
\end{multline}
One can finally introduce Majorana modes and obtain the final expression for our effective Hamiltonian:
\begin{align*}
\psi_{R/L}=\frac{\gamma^0_{R/L}+i \gamma^1_{R/L}}{\sqrt{2}}
\end{align*}
\begin{multline}
H=\sum\limits_{\sigma=0,1} \int dx \frac{(-i v_{F,+})}{2} \gamma^\sigma_R \partial dx \gamma^\sigma_R - \gamma^\sigma_L \partial dx \gamma^\sigma_L \\
+\int dx \frac{\pi(g_+ + \Delta^{(1)})}{\alpha} i \gamma^0_R \gamma^0_L + \frac{\pi(g_+ - \Delta^{(1)})}{\alpha} i \gamma^1_R \gamma^1_L.
\end{multline}
The effective model is consequently very simple: two massive Majorana fermions. A phase transition consequently occurs when one of the two masses vanishes, i.e $g_+ \pm \Delta^{(1)}=0$. At these two points, one of the Majorana wire is free while the other is massive. The transition is therefore an Ising transition instead of a Commensurate-Incommensurate transition, with a central charge $c=\frac{1}{2}$. We argue numerically that this picture is still valid when $\Delta$ and $g$ are of the same order by computing the central charge on the transition line. \\

 The transition between the $MI$-$F$ phase and the $4MF$ phase follows the same physics at $\mu=0$ by symmetry. Again, numerically the picture is still valid when one branches $\mu$.\\
 
 Finally the direct transition between the $MI$-$F$ phase and the Polarized phase is the simplest to describe, as it is apparent in the critical model presented in Section IVC. As explained in the previous Section, the effective model at large coupling is a Quantum Ising model in a Transverse Field. The critical model is consequently also the critical Ising model, with a central charge $c=\frac{1}{2}$.

\section{Phases at large chemical potential}\label{sec:DCI}

In this model, numerical  studies demonstrated the existence of a small gapless phase appearing far from half filling. While bosonization is a useful tool to study one-dimensional system at close to half-filling, the standard approach usually breaks down once one gets close to the bottom of the fermionic bands. In order to take into account the non-linearity of the free spectrum, one has to add supplementary terms such as $(\partial_x \phi)^3$ that prove very challenging to take into account \protect\cite{Haldane,Adilet}. To avoid such technicalities, we will reformulate our problem in an alternative and physical basis that will allow us to apply bosonization again. In this section, we present some analytical and numerical results on the properties of this gapless phase.

\subsection{Unraveling the ladder}

To find an effective model far from half-filling, we rewrite our fermionic wires to form two Majorana chains as discussed in Sec.~\ref{sec-kitaev} (see Eq. \ref{Eq:MajoranaWire}) and recombine them to form a single fermionic chain with a doubled number of lattice sites. Reinterpreting the surviving $U(1)$ symmetry  (rotation around the $y$ axis discussed in Sec.~\ref{sec-hubbard} as the conservation of fermionic charge in this new basis, we use bosonization to investigate physics close to Kitaev's critical point.\\
The additional index $\sigma$ specifies the wires. We place ourselves at $t=\Delta$ to clarify the physical picture \and pose $\delta \mu = \mu + 2t$. Our complete model can be rewritten in terms of the $\alpha$ operators, extending Eq.~\eqref{Eq:MajoranaWire},
\begin{multline}\label{eq:pairing}
H=-\frac{i \delta \mu}{2}\sum\limits_{j,\sigma} \alpha_{2j+1,\sigma} \alpha_{2j,\sigma}-it \sum\limits_{j,\sigma} \alpha_{j,\sigma} \alpha_{j+1,\sigma}
\\-\frac{g}{4}\sum\limits_{j,\sigma}\alpha_{2j,1} \alpha_{2j+1,1}\alpha_{2j,2} \alpha_{2j+1,2}.
\end{multline}

At the point $\delta\mu=0$ and $g=0$ the central charge at the transition is simply $c=1$, corresponding to two $c=1/2$ Majorana chains. We show below that the free fermion phase $DCI$  we find between the topological and the polarized phase is nothing but an extension of the critical point at $g=0$ with $c=2\times \frac{1}{2}=1$.\\

\begin{figure}
\def\svgwidth{\columnwidth}
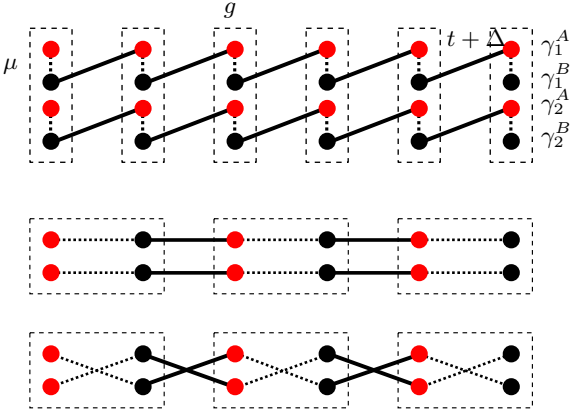
\caption{(Color online) Unravelling scheme to obtain the model far from half-filling. The initial fermions are split into two Majoranas. Each fermionic wire is then reorganized into a Majorana wire with an alternating hopping term. Finally, recombination of the Majorana wires into new fermions composed of a Majorana of each wire.}
\label{Fig:Unravel}
\end{figure}

We define new fermions mixing both wires, depicted in Figure \ref{Fig:Unravel},
\begin{subequations}\label{eq:def}
\begin{align}
d_{2j}&=\frac{(-1)^j}{2}(\alpha_{2j,1}+i\alpha_{2j,2}) \label{Eq:NewF1}\\
d_{2j+1}&=\frac{(-1)^{j+1}}{2}(\alpha_{2j+1,2}-i\alpha_{2j+1,1}). \label{Eq:NewF2}
\end{align}
\end{subequations}
and leading to the final Hamiltonian
\begin{multline}
H=-(2t-\frac{\delta \mu}{2}) \sum\limits_j (d^\dagger_j d_{j+1}+d^\dagger_{j+1} d_j)\\
-\frac{\delta \mu}{2} \sum\limits_j(-1)^j(d^\dagger_j d_{j+1}+d^\dagger_{j+1} d_j)\\
-\frac{g}{2} \sum\limits_j (1+(-1)^j)(n_j-\frac{1}{2})(n_{j+1}-\frac{1}{2}). \label{Eq:HNP}
\end{multline}

The even-odd site Majorana pairing of Eq.~\eqref{eq:pairing}  $\propto \delta \mu$ has been decomposed as an alternating sign term and an additional contribution to hopping. The alternating sign term favors dimerization depending on the sign of $\delta \mu$, just as discussed in Sec.~\ref{sec-kitaev} for a single Majorana chain. $g$ also separates into two contributions, one is alternating and the other is constant. The former also favors dimerization (this will appear more clearly below using Bosonization) and competes with $\delta \mu$, while the latter tries to impose a uniform charge distribution in competition with the two dimerization schemes.
\\

The boundary conditions in this basis will depend on those of the initial model. For OBC in the original model, it will also have OBC in this effective model. For PBC, it will either have PBC if $L$ is even or Anti-Periodic Boundary Conditions (APBC) if $L$ is odd. In the following, we consider $L$ to be even, but the conclusions will remain unaffected by this choice.\\

The definitions of Eqs.~\eqref{eq:def} reveal the hidden $U(1)$ symmetry discussed in Section~\ref{sec-hubbard} and consequently restore the conservation of the number of electrons in this basis. In the absence of chemical potential, we can apply an Abelian bosonization approach to study our model. We use the same convention as in Section \ref{subsec:boson}
and separate each fermion field into its left- and right-moving part:
\begin{equation}
d_{j} = e^{ik_F j} d_{1,j} + e^{-ik_F j}d_{-1,j}.
\end{equation}
\begin{equation}
d_{r,j}=\frac{U_{r}}{\sqrt{2 \pi \alpha}} e^{-i(r\phi_{j}-\theta_{j})}.
\end{equation}
We emphasize that there are no way of expressing linearly the bosonic field $(\phi, \theta)$ in terms of the usual charge and spin modes.

While most bosonized terms are standard, obtaining the correct contribution for the alternating part of $g$ is actually more challenging. One has to take special care and proceed to do the OPE of the term in order to get the correct expression (see for example Ref. \onlinecite{Orignac1998} where such terms are included to take into account disorder):
\begin{equation*}
(-1)^j n_j n_{j+1} \propto \partial_x \phi_j \sin(2\phi_{j+1}) \rightarrow \cos(2 \phi_j).
\end{equation*}

The final Hamiltonian is given by:
\begin{equation}\label{eq:boson}
H=\frac{\tilde{v}_F}{2 \pi}\left(\frac{1}{K} (\partial_x  \phi)^2 + K(\partial_x \theta)^2 \right)-g_\phi \cos(2\phi)
\end{equation}
with $\tilde{v}_F=(4t-\delta \mu)\sqrt{1-\frac{2g}{\pi (4t-\delta \mu)}}$, $g_\phi=(\frac{\delta \mu}{2\pi \alpha} + \frac{g}{2\pi^2 \alpha^2})$ and $K=\frac{1}{\sqrt{1-\frac{2g}{\pi (4t-\delta \mu)}}}$. 

Mapping our original two-channel model onto a single chain, of however doubled size, may seem to have reduced the number of fermionic modes leading to a halved central charge (from $c=2$ to $c=1$). It can be understood by noting that the critical point $\delta \mu=0$, $g=0$, corresponds to  the bottom of the band for the original fermions where linearization obviously breaks down and negative energy states disappear. In terms of the new fermions, the bosonized Hamiltonian of Eq.~\eqref{eq:boson} is obtained by discarding the operator $(-1)^j \partial_x \theta$. While this operator is irrelevant in the RG sense, it invalidates the bosonization procedure when too large. The discussion in this Section is therefore restricted to having $\delta \mu/t \ll 1$. \\

The dimension of $\cos(2\phi)$ is $K$. Consequently, as long as $K<2$, it is always relevant. The resulting ground state depends on the sign of $g_\phi$ and a quantum phase transition occurs at vanishing $g_\phi$. By continuity with the non-interacting case, $g=0$, we identify these two phases as two Majorana dimerized states translated by one site, corresponding to the standard and topological superconducting phases, see Sec.~\ref{sec-kitaev}, or polarized and $4MF$ phases in our model. The dimerization can also be discussed by rewriting the cosine term in terms of the original Majorana fermions
\begin{multline*}
- g_\phi \cos(2\phi)\propto g_\phi \sum\limits_j(-1)^j(d^\dagger_j d_{j+1}+d^\dagger_{j+1} d_j) \\
=\frac{i g_\phi}{2}\sum\limits_{j,\sigma}(\gamma_{A, j,\sigma}\gamma_{B,j,\sigma}-\gamma_{B,j+1,\sigma}\gamma_{A,j,\sigma}).
\end{multline*}
If $g_\phi$ is positive, the links between neighboring Majorana operators $\gamma_{B,j+1,\sigma}\gamma_{A,j,\sigma}$ is favored, forming the topological phase where the Majorana fermions $\gamma_{B,1,\sigma}$ and $\gamma_{A,L,\sigma}$ on the extremities are left unpaired. In the opposite case of a negative $g_\phi$ the links between same site Majoranas $\gamma_{A, j,\sigma}\gamma_{B,j,\sigma}$ is favored, making the original fermions the good quasi-particles, i.e we find the trivial (polarized) phase.\\
When $K>2$, the cosine term stops being relevant: a gapless phase opens around the line $g_\phi=0$. This phase is a $c=1$ Luttinger Liquid in the language of the $d$ fermions and therefore an extension of the critical point at $g=0$. 

The opening of this DCI phase can be understood in the following way: while $\delta \mu$ and the alternating part of $g$ tend to form two types of contradictory dimerizations, the constant part of $g$ opposes the two. When $g$ is large enough, it prevents any of them from occurring. 

\subsection{Large $g$ model}\label{subsec-largeg}

An interesting limit to study is the behavior of the DCI when $g\rightarrow +\infty$. Indeed, from bosonization, one expects it to survive at infinite coupling, at the vicinity of the point $\mu=\frac{\pm g}{2}$. We focus in this section on $\mu=-\frac{g}{2}+\delta \mu$, with $\delta \mu=O(t)$. At this point, there is either zero or one fermion on each rung of the ladder. It is possible to derive an effective model similar to the $t-J$ model for Hubbard, but we will be interested in the model at $0^\text{th}$ order in $g$.\\
We define dressed fermions $\tilde{c}_{j,\sigma} = c_{j,\sigma}(1-n_{j, -\sigma})$, where $n_{j, -\sigma}$ is the number operator at site $j$ on the wire $2$ if $\sigma=1$ (and $1$ if $\sigma=2$). These dressed fermions follow a different algebra than usual but allow for a simple writing of the Hamiltonian (and a direct implementation for numerics):
\begin{align*}
\{\bar{c}_{i, \sigma}, \bar{c}_{j, \sigma'}\} &= 0~~~~\{\bar{c}_{i, \sigma}, \bar{c}^\dagger_{j, \sigma}\}=\delta_{i,j}(1-n_{j,-\sigma})\\
\{\bar{c}_{i, \sigma}, \bar{c}^\dagger_{j, -\sigma}\}&=\delta_{i,j}c^\dagger_{j,-\sigma}c_{j,\sigma}=\delta_{i,j}\bar{c}^\dagger_{j,-\sigma}\bar{c}_{j,\sigma}
\end{align*}
\begin{multline}
H=-\delta\mu \sum\limits_{j, \sigma} c_{j,\sigma}^\dagger c_{j,\sigma} -t \sum\limits_{j, \sigma} \left(\tilde{c}_{j+1,\sigma}^\dagger \tilde{c}_{j,\sigma}+ \tilde{c}_{j,\sigma}^\dagger \tilde{c}_{j+1,\sigma} \right)\\
+\Delta \sum\limits_{j, \sigma} \left(\tilde{c}_{j,\sigma}^\dagger \tilde{c}^\dagger_{j+1,\sigma}+ \tilde{c}_{j+1,\sigma} \tilde{c}_{j,\sigma} \right). \label{Eq:HLG}
\end{multline}

The definition as dressed fermions comes naturally from the restriction to a three dimensional subspace on each site. It is then just as natural to try to find an equivalent system replacing the fermions under constraints by spin one. It is possible to construct a Jordan-Wigner like transformation that verifies the previous algebra, with $S^x$, $S^y$ and $S^z$ the usual spin-1 operators.
\begin{align}
\bar{c}^\dagger_{j, \uparrow}&=S_j^z S_j^+ e^{i\pi \sum\limits_{k<j} (S^z_k)^2} \nonumber \\
\bar{c}^\dagger_{j, \downarrow}&=-S_j^z S_j^- e^{i\pi \sum\limits_{k<j} (S^z_k)^2}.
\end{align}
After a bit of algebra, one has an alternative expression for the Hamiltonian:
\begin{multline}
H=-\mu (S_z)^2+\frac{\Delta-t}{2}\{S^x_jS^x_{j+1}, S^z_jS^z_{j+1}\}\\
-\frac{\Delta+t}{2}\{S^y_jS^y_{j+1}, S^z_j S^z_{j+1}\} -\frac{t}{2}(S^x_j S^x_{j+1}+ S^y_j S^y_{j+1}).
\end{multline}

Both these models are not simply solvable, but they can be efficiently treated by numerical means. In contrast to the (solvable) spin-1 chain models\protect\cite{Haldane1983, AKLT1987, Takhtajan1982, Babujian1982}, this Hamiltonian is not easily solvable and the Poisson brackets do not simplify easily. We will therefore resort to a numerical analysis below showing that the DCI phase becomes well visible in the phase diagram in the limit of large interactions.

\subsection{Numerical probes}
To confirm the analytical predictions of the previous bosonization, we studied several numerical observables using different models.

\subsubsection{Existence of the $DCI$ phase}

While bosonization affirms the existence of the phase for $K>2$, there exist well-known examples where there is a limiting value for $K$ that is not trivially detectable but appears when one exactly solves the model. A canonical example is, for example, the limit $\frac{1}{2}<K<2$ in the Hubbard model. If we had such a limit, there would only have a transition line for all $g$. A first numerical approach is to work at fixed $g$ and try to interpolate the boundaries of our supplementary phase in the thermodynamic limit. To determine the boundary of the phase, one can consider either the closing of the gaps or the peak in central charge, as both neighboring phases are gapped. Nevertheless, the most visible numerical marker of the phase will be the degeneracy of the ground state in the case of PBC. In that case, the ground state is doubly degenerate, and the ground states have different fermionic parities: (odd, even) and (even, odd). This spontaneous breakdown of the symmetry between the wires is allowed, as we break only a discrete symmetry, similarly to what happens in the two Mott phases. Mean-field computations allow to intuitively understand this degeneracy. In analogous fermion models (for example the XXZ model), such a degeneracy was observed for PBC when the length of the wire is a multiple of 4 sites\protect\cite{sutherland2004beautiful, Loss1992}.

\begin{figure}
\begin{center}
\includegraphics[width=0.5\textwidth]{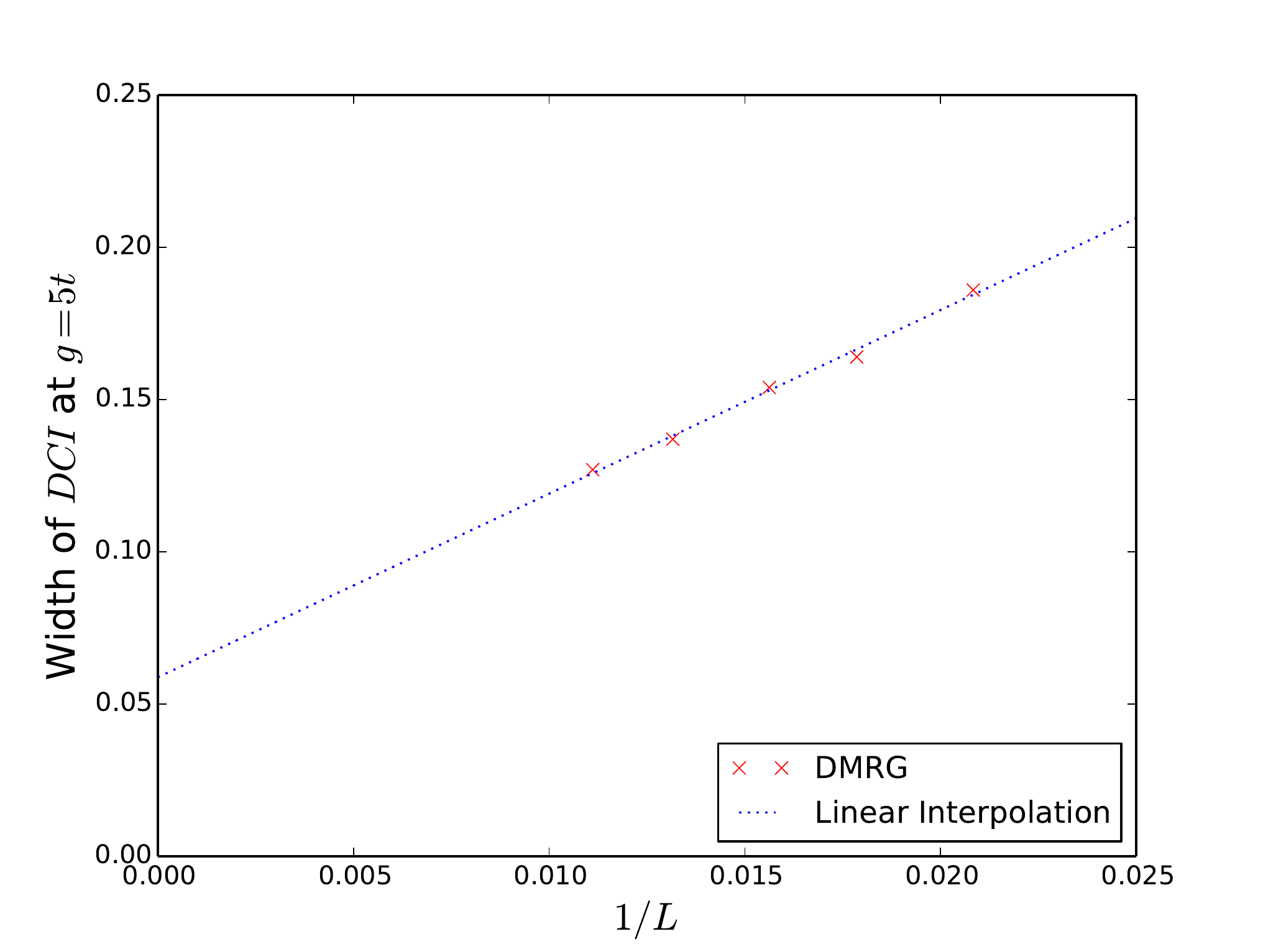}
\end{center}
\caption{(Color online) Linear regression of the width of the $DCI$ phase for $g=5t$ and $\Delta=t$ based on the analysis of the energy spectrum. This width converges towards a finite non-zero value of order $0.06 t$.}
\label{fig:WNP}
\end{figure}

Figure \ref{fig:WNP} presents the results of such a scaling analysis for $\Delta=t$ and $g=5t$. The width converges towards a finite value $0.06t$.

\begin{figure}
\begin{center}
\includegraphics[width=0.5\textwidth]{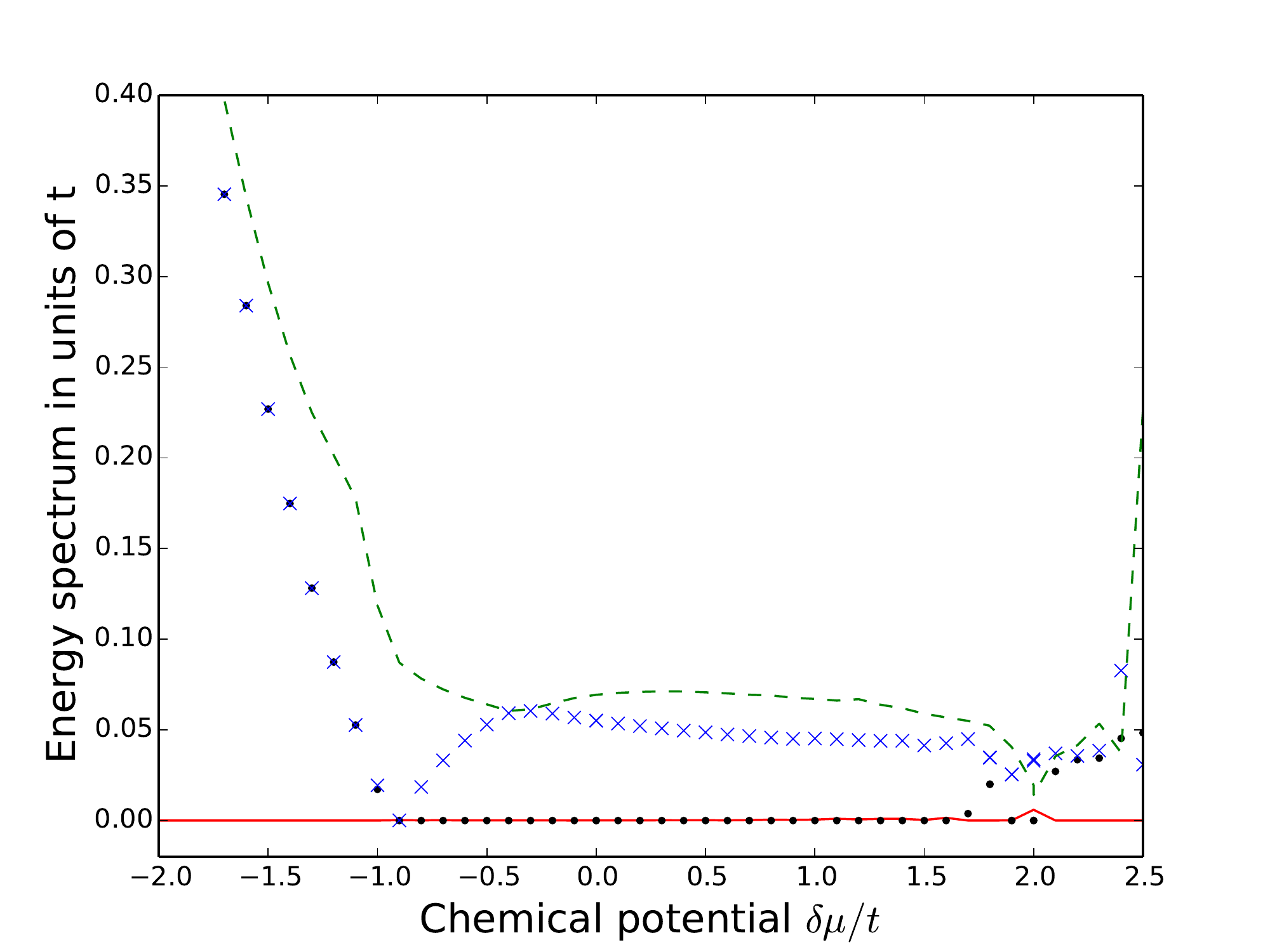}
\end{center}
\caption{(Color online) First four levels of the energy spectrum at $g=+\infty$ for a 68-site ladder at $\Delta=t$ from DMRG with PBC. The visible double degeneracy of the central phase reveals the survival of the $DCI$ phase. Scaling analysis and entanglement entropy confirms its gaplessness.}
\label{fig:LGNP}
\end{figure}

\begin{figure}
\begin{center}
\includegraphics[width=0.5\textwidth]{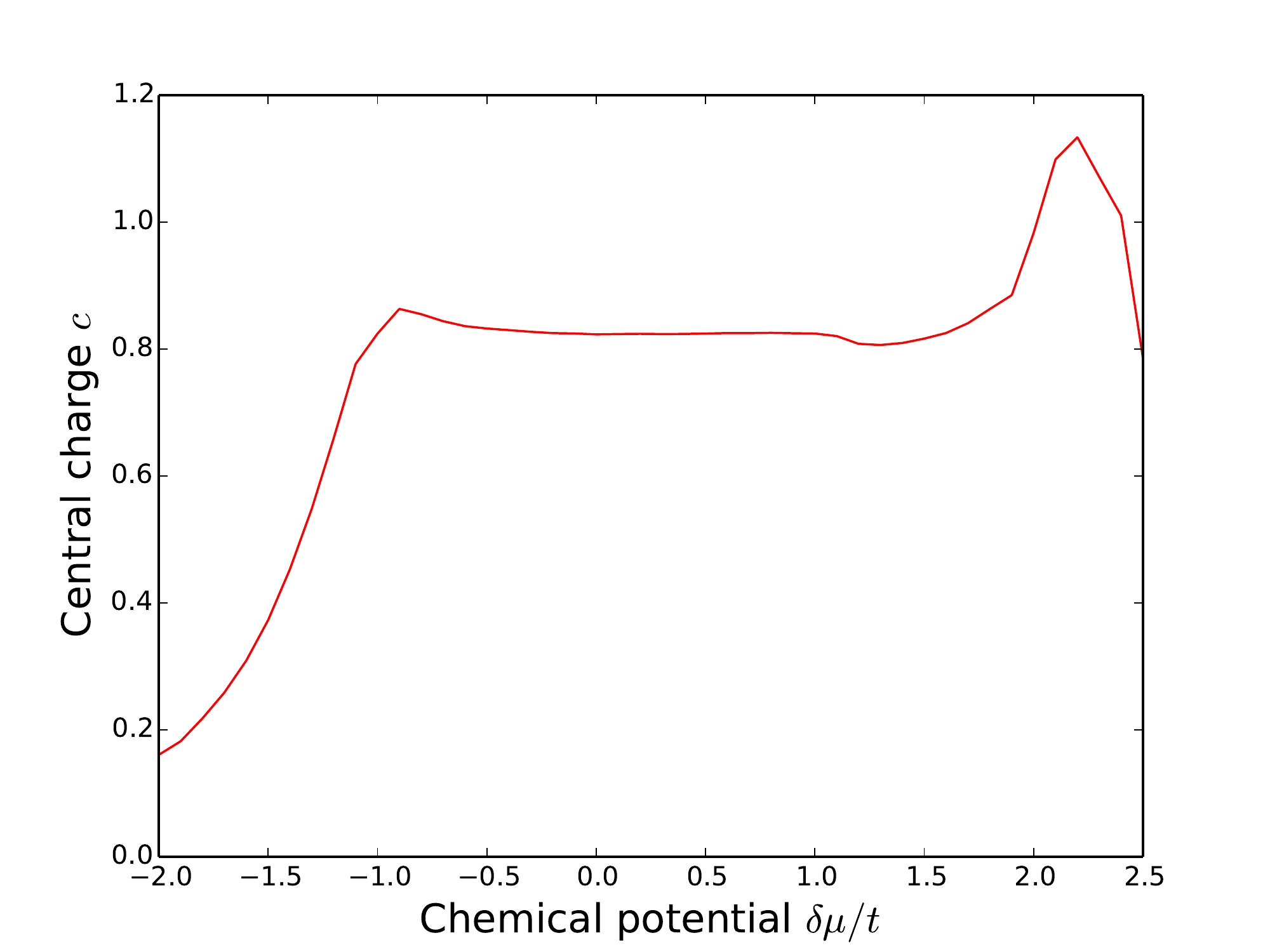}
\end{center}
\caption{(Color online) Central charge obtained from entanglement entropy for a 68-site ladder at $\Delta=t$ and $g=+\infty$ from DMRG with PBC, using the model given in Eq \ref{Eq:HLG}. The results are in qualitative agreement with our theoretical prediction and the results at finite $g$ in Fig \ref{fig:CCNP}.}
\label{fig:LGCC}
\end{figure}

As another element of answer, one can consider the limiting models for $g\rightarrow + \infty$ we previously derived. Figure \ref{fig:LGNP} presents the first four levels of the Hamiltonian \ref{Eq:HLG} for a range of renormalized chemical potential. The double degeneracy is symptomatic of the $DCI$ phase. The absence of a gap in this phase is also confirmed by both scaling analysis and entanglement entropy. Figure \ref{fig:LGCC} presents the central charge computed from the entanglement entropy of this model, in good agreement with the results at finite $g$ and our theoretical predictions. 

We should also discuss the width of the phase in terms of the chemical potential at $g=+\infty$. While it was extremely limited at finite $g$, it is now of order $3t$ for the system considered in Fig \ref{fig:LGNP} and \ref{fig:LGCC}. On the other hand, at infinite $g$, the $4MF$ phase has disappeared. 

\subsubsection{Properties of the $DCI$ phase}

First we focus on numerical results obtained using our original model. As in Section IV, we will use different numerical markers: entanglement entropy and spectrum, spectrum with PBS and bipartite charge fluctuations.

\begin{figure}
\begin{center}
\includegraphics[width=0.5\textwidth]{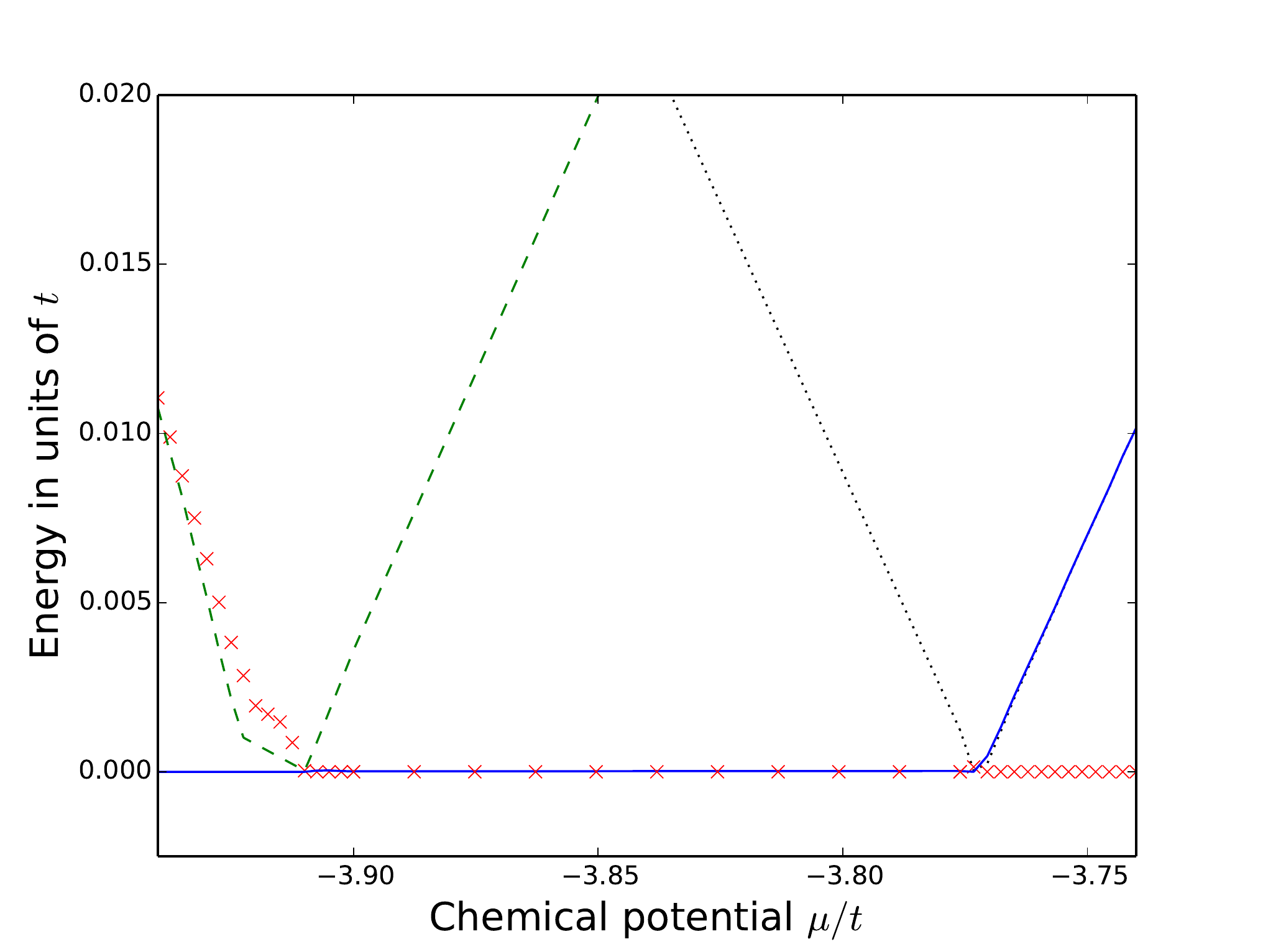}
\end{center}
\caption{(Color online) First four levels of the energy spectrum of the periodic ladder for $g=5t$ and $\Delta=t$, as a function of the chemical potential for a $76$-site wire. On the left, the system is in the polarized phase, on the right it is in the $4FM$ phase. The double degeneracy of the ground state clearly reveals the $DCI$ phase. Symmetry analysis reveals that the parity of the ground states follow the stated rules (even-even in polarized, odd-odd in $4FM$ and (even-odd, odd-even) in $DCI$.}
\label{fig:SpecNP}
\end{figure}

\begin{figure}
\begin{center}
\includegraphics[width=0.5\textwidth]{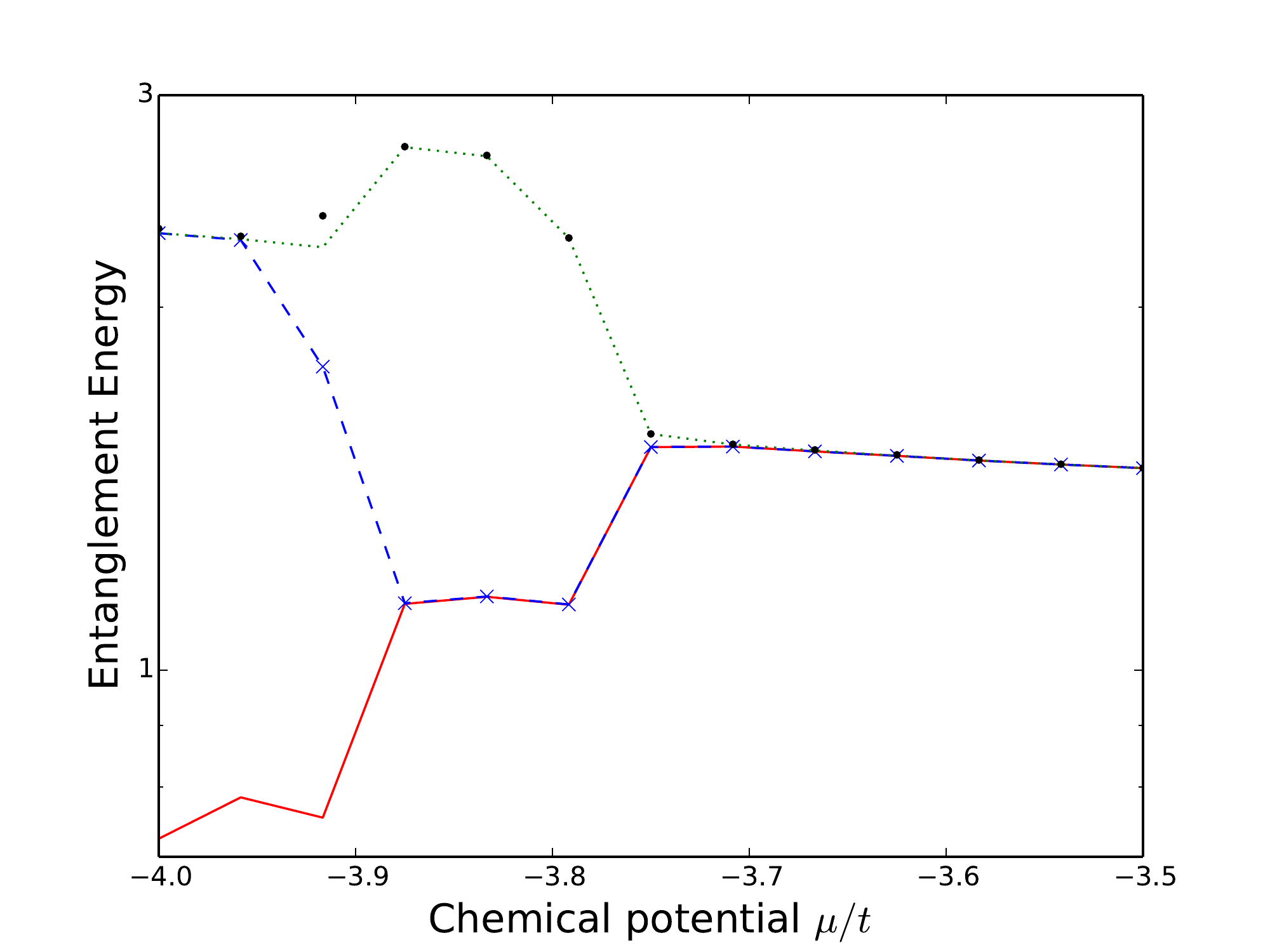}
\end{center}
\caption{(Color online)Entanglement Spectrum of the periodic ladder for $g=5t$ and $\Delta=t$, as a function of the chemical potential for a $90$-site wire. On the left, the system is in the polarized phase, on the right it is in the $4FM$ phase. We observe a progressive lifting of the degeneracy from $4$ in the $4FM$ to $2$ in the $DCI$ and finally no degeneracy in the Polarized phase.}
\label{fig:EGNP}
\end{figure}

\begin{figure}
\begin{center}
\includegraphics[width=0.5\textwidth]{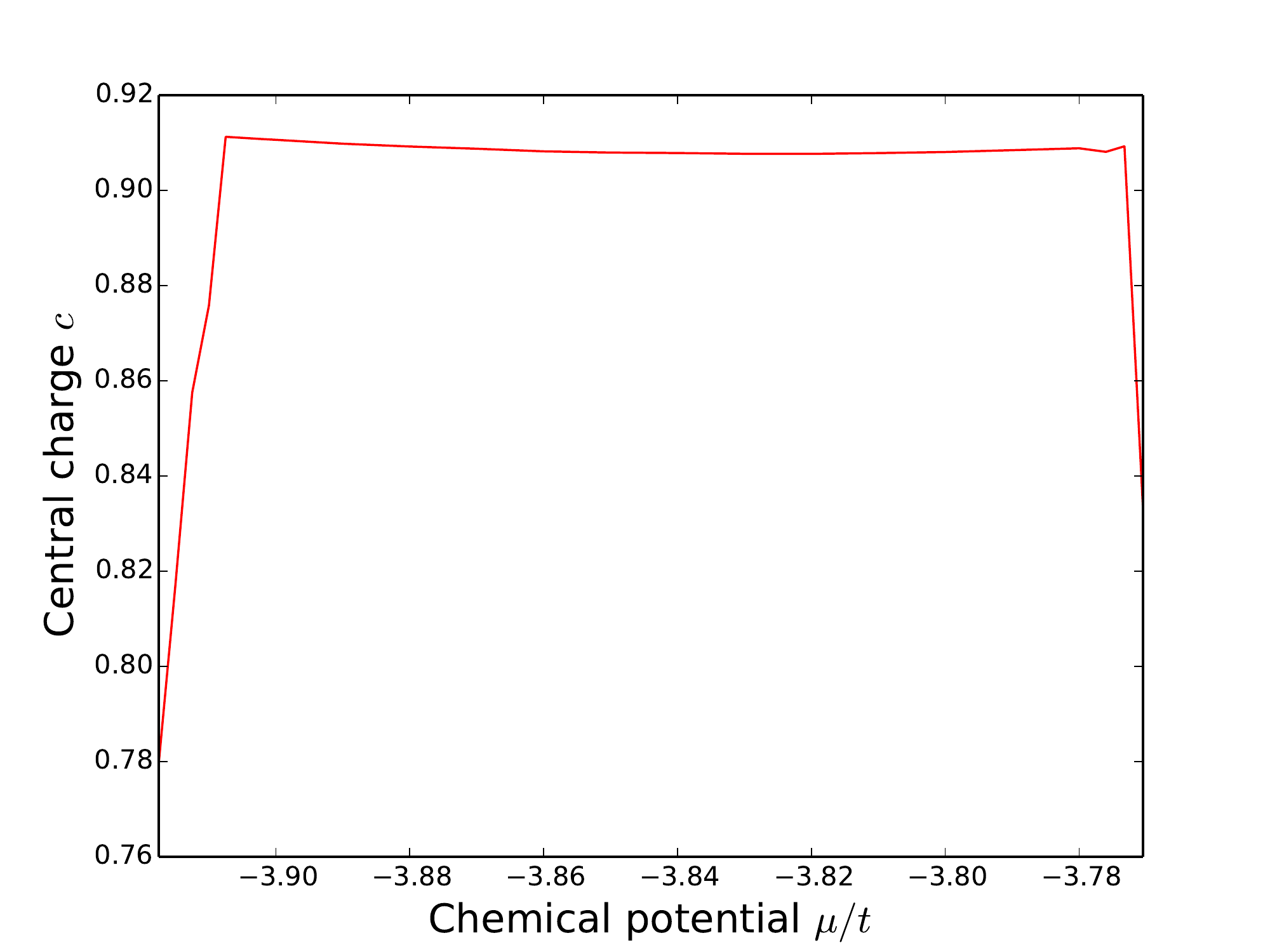}
\end{center}
\caption{(Color online) Central charge extracted from a fit of the entanglement entropy of the periodic ladder for $g=5t$ and $\Delta=t=1$, as a function of the chemical potential for a $76$-site wire. }
\label{fig:CCNP}
\end{figure}

The spectrum for the ladder with PBC is given in Figure \ref{fig:SpecNP} indeed, we observe for positive and large values of $g$ a gapless phase with an accidental two-fold degeneracy that disappear with OBC. Their parity follow the rules stated in the previous section. Scaling of the finite-sized gap in this phase reveal its gaplessness. We observe also a corresponding two-fold degeneracy in the entanglement spectrum, compared with four-fold for the SPT phase and no degeneracy for the polarized phase in Figure \ref{fig:EGNP}. Finally, one can compute the central charge from the entanglement entropy: it confirms the gaplessness of the phase and validate our model with the extraction of a charge close to unity in the $DCI$ phase (Figure \ref{fig:CCNP}).

\begin{figure}
\begin{center}
\includegraphics[width=0.5\textwidth]{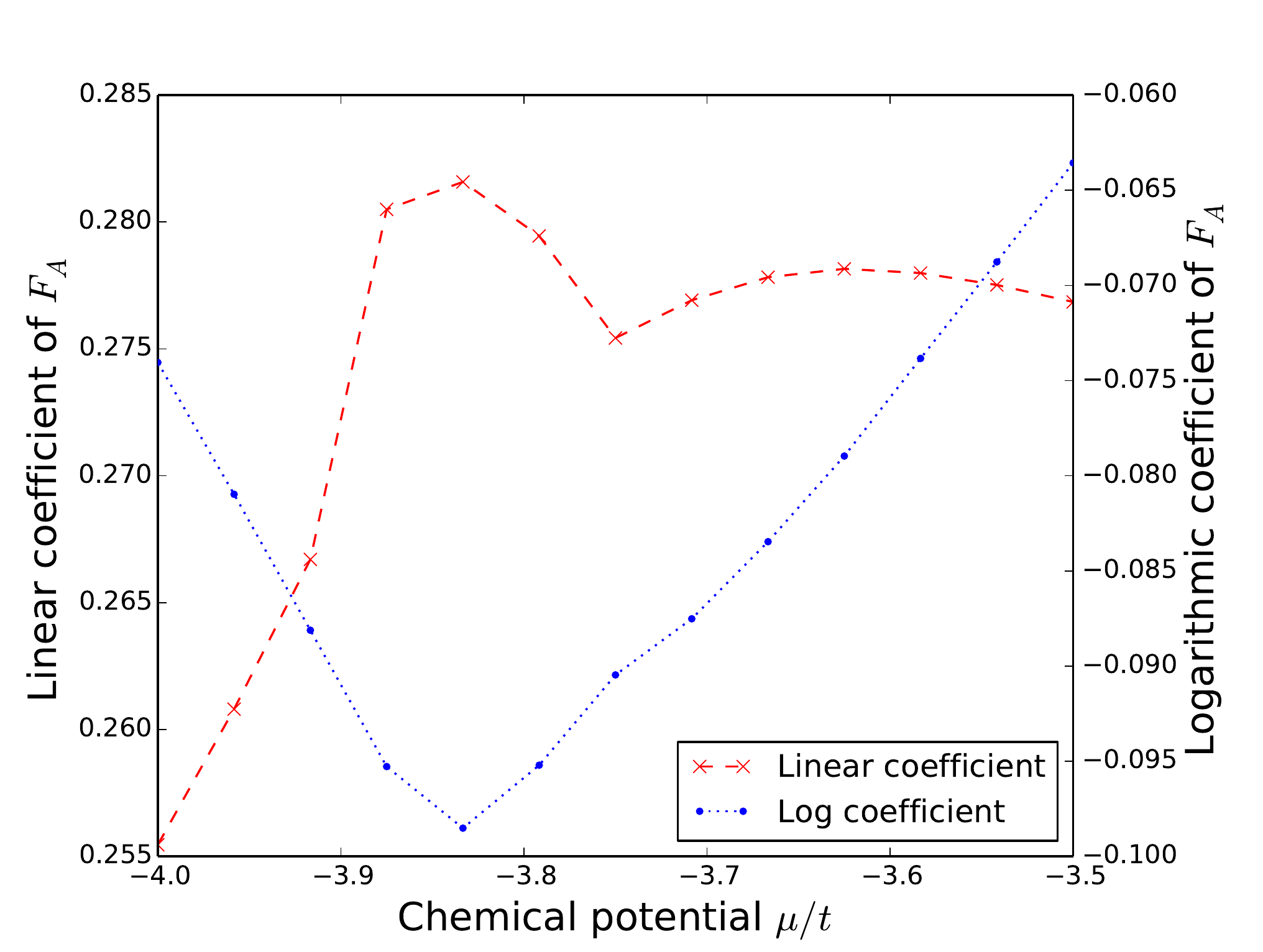}
\end{center}
\caption{(Color online) Logarithmic contribution to the bipartite charge fluctuations in the first wire in the $DCI$ phase for a $90$-site wire at $g=5t$ and $\mu=-3.85t$. The pic corresponds to the middle of this phase.}
\label{fig:FANP}
\end{figure}

The bipartite charge fluctuations at the transition strongly depend both on the critical theory and on how the density can be expressed in the critical language. Two wires both at the critical $c=\frac{1}{2}$ transition will present negative logarithmic fluctuations while one can unite the two critical models to form a simple bosonic $c=1$ model, whose associated density presents positive logarithmic contributions. The main reason behind numerically studying the bipartite fluctuations in our model is a verification that there exists no simple expression of the free mode of the $c=1$ phase in terms of the charge and spin modes, showing that the $DCI$ phase is indeed an expansion of the critical point at $g=0$. There are nonetheless no reason why the logarithmic contribution at $g\neq 0$ would be the same that at the non interacting point. Figure \ref{fig:FANP} presents the behavior of the logarithmic contribution around the $DCI$ phase for a $90$-site wire at $g=5t$ and $\Delta=t$.

\section{Conclusion}\label{sec:conclusion}

To summarize, we have studied the phase diagram and entanglement properties of two interacting Kitaev chains. The model interpolates between two standard models of low-energy physics, the superconducting Kitaev model and the
Hubbard model. At weak to moderate interactions, we have shown through bosonization and renormalization group arguments that the superconducting phase with 4 Majoranas at the extremities of the ladder is stable. The presence of these Majorana bound states are also detected numerically through the entanglement spectrum and through the four-fold degeneracy of the ground state (with open boundaries). For large interactions, Mott phases occur by analogy with the Hubbard model. The Cooper pairing term breaks the SU(2) spin symmetry and results in an Ising type order. This engenders orbital currents already without the application of a net magnetic flux in the system.  The competition between the Cooper pair channel and the Coulomb term also reveals an exotic phase in the phase diagram, called the DCI phase, in the vicinity of the polarized phases occurring at large chemical potentials. Using exact mappings (onto  a dimerized wire point of view and a dressed fermion or equivalently a spin-1 chain representation), the concept of bipartite fluctuations and numerical computations, we have found that this phase can be identified to two critical Ising models, as a reminiscence of the quantum critical point in the Kitaev model. We have also justified that this phase occurs at a critical value of the Coulomb interaction between chains. The double degeneracy of the ground state for PBC can be justified from a mean-field point of view.  It is also relevant to note that the DCI phase (as well as the critical point of the Kitaev model) can be identified through negative sub-leading logarithmic corrections in the bipartite fluctuations, that we have computed in the non-interacting case.

Such an interacting Kitaev ladder could be realized with current technology either using quantum wires with spin-orbit coupling or through two quantum Ising chains in Josephson junction systems or in ultra-cold atoms. The strong $g$ limit could be achieved with two-coupled quantum Ising chains through the Hamiltonian (3). With superconducting wires, the capacitive coupling between chains could be enhanced by inserting an insulating material between the two quantum wires. We also note that the DCI phase is related to two chains of Majorana modes (present at each sites) and could be potentially relevant for quantum information purposes \protect\cite{Terhal2012}. The effects of additional interactions and coupling terms  could be addressed in a future work.

\section*{ACKNOWLEDGEMENTS}

This work has benefited from useful discussions with I. Affleck, S. Diehl, J.-N. Fuchs, M. Franz, A. Jouan, L. Henriet, Ph. Lecheminant, R. Lutchyn, A. Petrescu, S. Rachel, H. F. Song and at the CIFAR meetings in Canada, workshops in KITP Santa Barbara, Trieste and London, and at a DFG meeting in M\" unich. We acknowledge support from the PALM Labex, Paris-Saclay, Grant No. ANR-10-LABX-0039.

\begin{appendix}
\section{Bogoliubov formalism and observables in Kitaev wire \label{app:Bogo}}
In this Appendix, we introduce Bogoliubov formalism for Kitaev wire. It allows for an exact solution of the model for periodic boundary conditions (PBC), and the computation of different observables. We show that the density is not a good order parameter and introduce the effective pairing, related to BCS gap. The effective pairing is a good order parameter for the topological phase transition in Kitaev model, but is affected by finite size effects.

\subsection{Bogoliubov quasi-particles formalism}
In the case of periodic boundary conditions (PBC), the quadratic Hamiltonian is easily diagonalized by a Bogoliubov transform. To avoid any possible confusion, the Fourier transform convention we use is the following: $c_k = \frac{1}{\sqrt{L}}\sum\limits_{j=1}^L e^{-ikj}c_j$.

We define $\Psi_{k}^\dagger=(c^\dagger_k, c_{-k})$. Forgetting constant terms, we can therefore write the Hamiltonian as :
$$H_K\{c \}=\frac{1}{2} \sum\limits_{k} \Psi_{k}^\dagger h(k) \Psi_{k},$$
with 
$$h(k) = \begin{pmatrix}
-\mu -2t \cos k & 2i\Delta \sin k \\
-2i\Delta \sin k & \mu +2t \cos k
\end{pmatrix}.$$

We define the angle $\theta_k$ by :
\begin{align*}
\cos(\theta_k)&=\frac{-\mu - 2t \cos k}{\sqrt{(\mu + 2t\cos k)^2 + 4\Delta^2 \sin^2 k}}\\
\sin(\theta_k) &= \frac{2\Delta \sin k}{\sqrt{(\mu + 2t\cos k)^2 + 4\Delta^2 \sin^2 k}}.
\end{align*}
We introduce the Bogoliubov quasi-particle operators $\eta_{k} = \cos(\theta_k/2) c_k + i \sin(\theta_k/2) c_{-k}^\dagger$ that diagonalize the Hamiltonian and $E_k=\sqrt{(\mu+2t\cos k)^2+4 \Delta^2 \sin^2 k}$.
$$H_K\{c\}=\sum\limits_{k} E_k \eta^\dagger_{k} \eta_{k}.$$

Noting $\Ket{0}_c$ the empty state for the $c$ fermions, the BCS ground state is given by:
\begin{multline}
\Ket{\text{BCS}}= ((\delta_{\mu <-2t} + (1-\delta_{\mu <-2t})c^\dagger_0)((\delta_{\mu <2t} + (1-\delta_{\mu <2t})c^\dagger_\pi)\\
\prod\limits_{k>0}^{k<\pi}\left(\cos(\theta_k/2)+i\sin(\theta_k/2)c^\dagger_k c^\dagger_{-k}\right) \Ket{0}_c.
\label{Eq:BCSGS}
\end{multline}
From the bulk-edge correspondence, we expect a closure of the gap when there is a quantum phase transition. Solving $E_k=0$, we find the expected phase diagram. 
Rewriting the original operators as a function of the $\gamma_k$, we can deduce the following average in the BCS ground state:
\begin{align*}
\Braket{c^\dagger_k c_q}&= \delta_{q,k}\sin(\theta_k/2)^2\\
\Braket{c^\dagger_k c^\dagger_q} &= \delta_{q, -k} \frac{i}{2} \sin(\theta_k).
\end{align*}
\subsection{Observables}\label{appsub:Observables}
We focus below on possible order parameters in Kitaev wires, in order to be able to find additional numerical indicators of the topological phase.
 
\begin{figure}
\begin{center}
\includegraphics[width=0.45\textwidth]{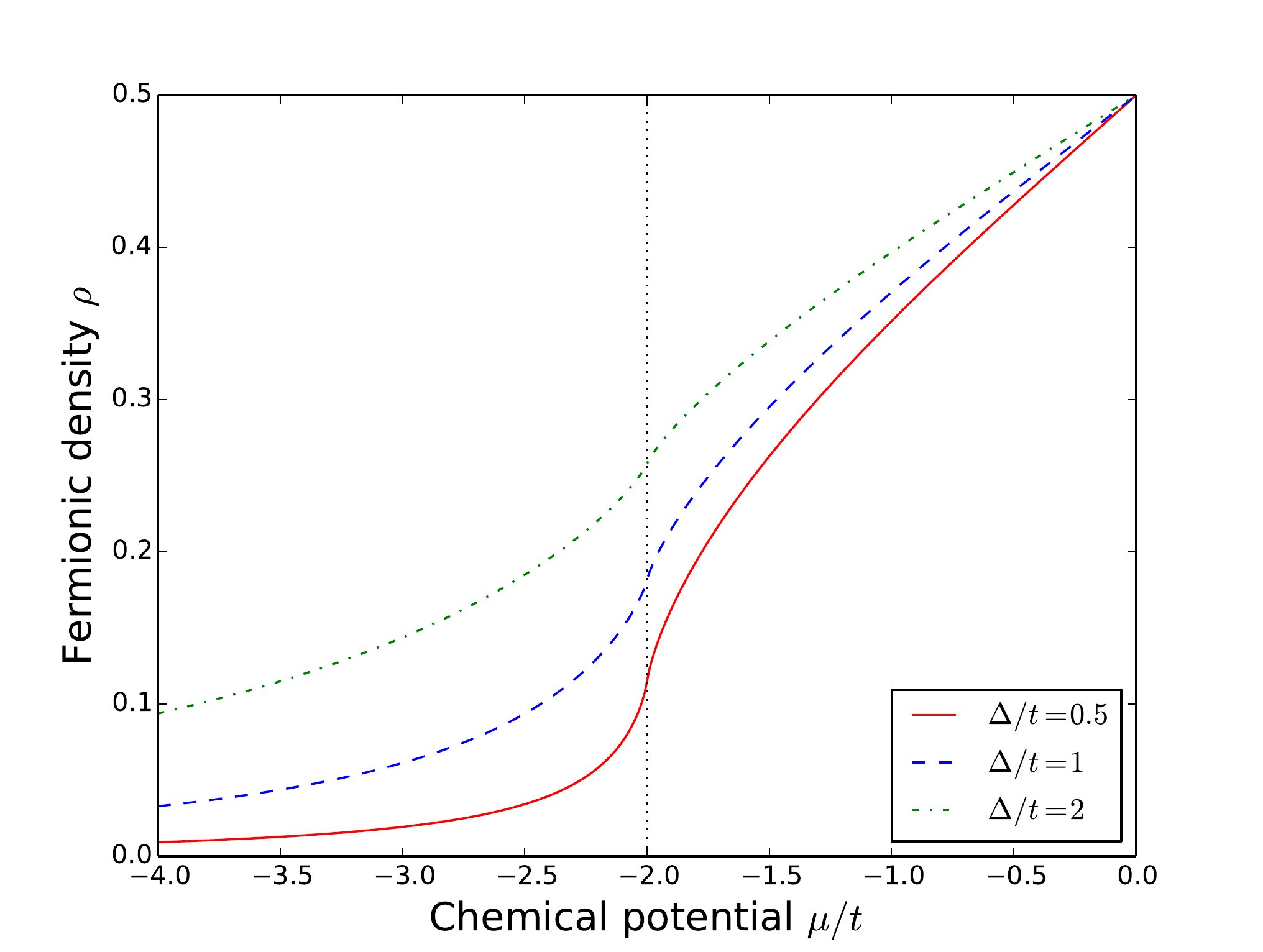}
\end{center}
\caption{(Color online) Fermionic density as a function of $\mu$ (in units of $t$) for the non-interacting Kitaev wire for different values of $\Delta$ (in red, $\Delta=2t$, in green $\Delta=t$, in blue $\Delta=0.5t$).}
\label{fig:DensNoInt}
\end{figure}

Correlation functions and order parameters have been considered a long time ago for the Ising model in a transverse field\protect\cite{pfeuty1970one}. The fermionic density in the wires without interaction is given by the following expression:
\begin{align}
\rho(\mu,t,\Delta)-\frac{1}{2}&=\frac{1}{L} \sum\limits_{j=1}^L \left(\Braket{c^\dagger_j c_j}-\frac{1}{2}\right)=\frac{1}{L} \sum\limits _{j=1}^L \Braket{\sigma^z_j}  \nonumber \\
&=-\int\limits_0^\pi \frac{dk}{4\pi}\frac{-\mu-2t\cos k}{\sqrt{(\mu+2t\cos k)^2+4\Delta^2\sin^2 k}}.
\end{align}
This integral is a well-known elliptical form that has been thoroughly investigated in the transverse Ising model. It corresponds to the magnetization of the Quantum Ising model. Figure \ref{fig:DensNoInt} presents the density of electrons in the wire as a function of the chemical potential for several values of $\Delta$. As it is continuous at the transition, it is not a good order parameter. It has nonetheless an analytical form on the transition line:

\begin{eqnarray}
\rho(2t,t,\Delta)-\frac{1}{2}&=&\left\lbrace \begin{array}{ll}
\frac{1}{\pi} \frac{\arcsin(\sqrt{1-\Delta^2/t^2})}{\sqrt{1-\Delta^2/t^2}} & \text{if $|\frac{\Delta}{t}|\le 1$}\\
\frac{1}{\pi} \frac{\text{argsh}(\sqrt{\Delta^2/t^2-1})}{\sqrt{\Delta^2/t^2-1}} & \text{else} \\
\end{array} \right.
\label{Eq:TransDens}
\end{eqnarray}

Another conventional order parameter for the Quantum Ising model is the magnetization in the $Y$-direction $\Braket{\sigma_y}$ (with our conventions) for a ferromagnetic model. For the spin model, the transition is of second order, and $\Braket{\sigma_y}$ as a critical exponent $\beta=\frac{1}{8}$ at the transition. For the fermionic model, it is neither natural nor convenient: due to the string, it is complex to evaluate and it breaks the charge parity as its expression includes an odd number of fermion operators: it is consequently zero in the ground state. Instead, we present another related quantity that can be easily computed. It is the analogue to the superconducting gap of the BCS theory:
\begin{equation}
\Delta_{\text{BCS}}=\frac{1}{iL}\sum\limits_{k=0}^\pi \Braket{c^\dagger_k c^\dagger_{-k}},
\end{equation}
where the sum carries on the positive discrete momenta on the lattice. We emphasize that $\Delta_{\text{BCS}}$ is not the gap of our system, but only quantify the effective pairing of our theory. In the thermodynamic limit, one can express it as another elliptical integral that is solvable
\begin{equation}
\Delta_{\text{BCS}}=\int\limits_0^\pi \frac{dk}{2\pi} \frac{\Delta \sin k}{\sqrt{(\mu+2t\cos k)^2+4\Delta^2\sin^2 k}}.
\end{equation}
After some cumbersome algebra, one can obtain an exact analytical expression of the effective pairing. To simplify expressions, in the two following results we use the notations $\lambda = \frac{\mu}{2t}$ and express $\Delta$ in unit of $t$.
\begin{equation}
\Delta_{\text{BCS}}= \left\lbrace \begin{array}{ll}
\frac{\Delta}{4\pi\sqrt{1-\Delta^2}} \log \left(\frac{\sqrt{1-\Delta^2}|\lambda+1| + \lambda + (1-\Delta^2)}{\sqrt{1-\Delta^2}|\lambda-1| + \lambda - (1-\Delta^2)}\right) & \text{if $\Delta<1$} \\
\frac{1}{4\pi\lambda} (|\lambda+1|-|\lambda-1|] & \text{if $\Delta=1$} \\
\frac{\Delta}{4\pi\sqrt{\Delta^2-1}}\left(\arctan \left(\frac{\sqrt{\Delta^2-1}|\lambda+1|}{\lambda + (1-\Delta^2)}\right)\right. & \\ 
 ~~~~\left. -\arctan \left(\frac{\sqrt{\Delta^2-1}|\lambda-1|}{\lambda- (1-\Delta^2)}\right)+\pi\right) & \text{if $\Delta>1$}.
\end{array}\right.
\end{equation}
From the expression, it is easy to see that the effective pairing is constant inside the whole topological phase, and quickly decrease outside of it. The critical exponent is here simply $1$. Indeed, from the precedent expressions, we obtain, for all $\mu \in [-2t, 2t]$:
\begin{equation}
\Delta_{\text{BCS}}= \left\lbrace \begin{array}{l}
\frac{\Delta}{4\pi\sqrt{1-\Delta^2}} \log\left(\frac{1+\sqrt{1-\Delta^2}}{1-\sqrt{1-\Delta^2}}\right) \quad \text{ if $\Delta<1$} \\
\frac{1}{2\pi} \quad \qquad \qquad \qquad \qquad \qquad \text{if $\Delta=1$} \\
\frac{\Delta}{4\pi\sqrt{\Delta^2-1}} \left(\arctan \left(\frac{2\sqrt{\Delta^2-1}}{2-\Delta^2} +\delta_{\Delta>\sqrt{2}}\pi\right) \right).
\end{array}\right.
\end{equation}

\begin{figure}
\includegraphics[width=0.45\textwidth]{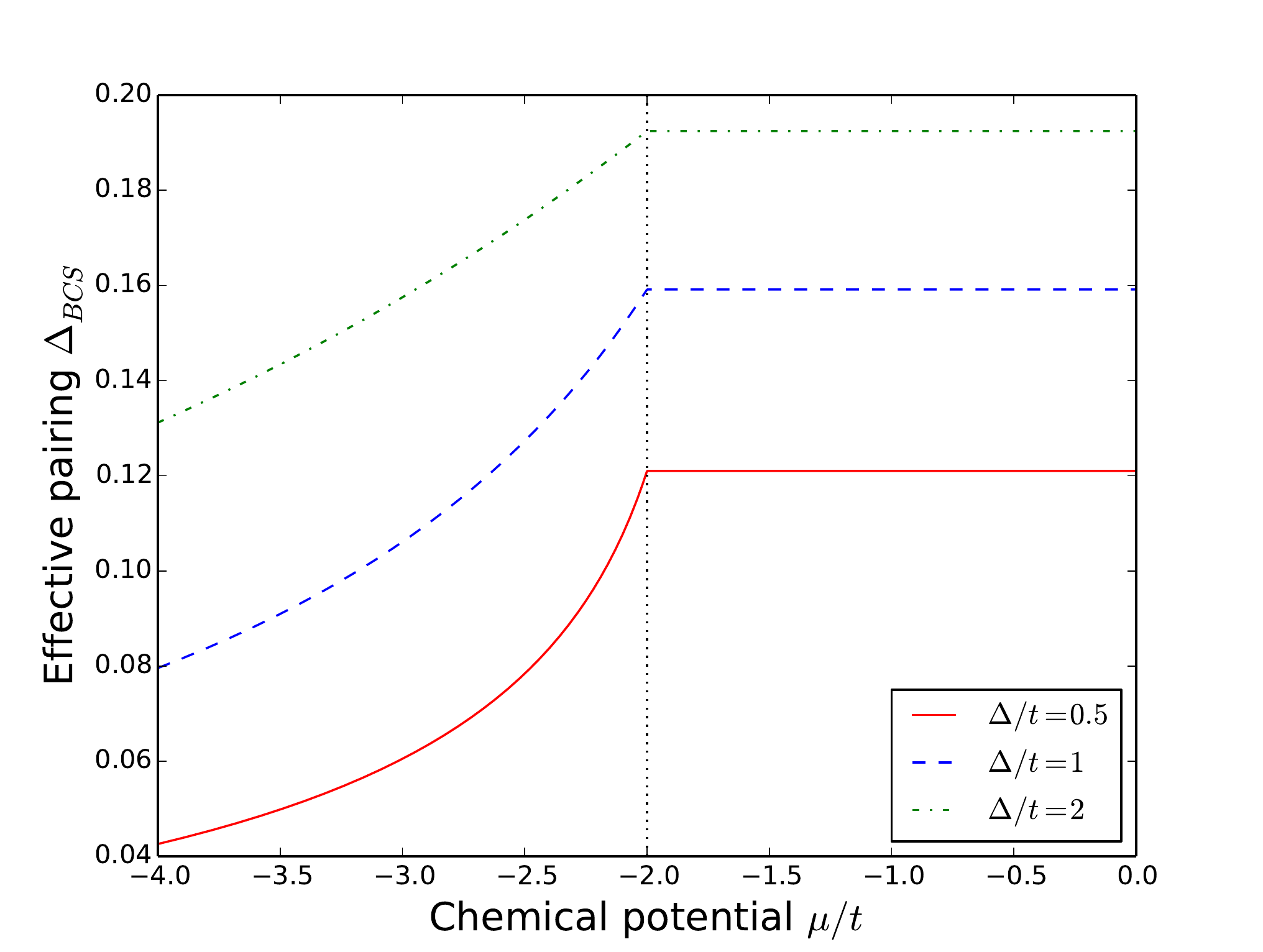}
\caption{(Color online) Effective pairing as a function of $\mu$ (in units of $t$) for non-interacting Kitaev wire for different values of $\Delta$ (in red, $\Delta=2t$, in green $\Delta=t$, in blue $\Delta=2t$). It serves as a good order parameter in the thermodynamic limit as it is constant in the topological phase.}
\label{fig:EffPaiNoInt}
\end{figure}

Figure \ref{fig:EffPaiNoInt} presents the effective pairing of the Kitaev model. As announced, the transition between the topological and the non topological phase is marked by a change in behavior of the effective pairing in the wire. As a marker in numerical studies, it is nonetheless limited by finite size effects leading to blurring oscillations. We consequently will be interested in another way to characterize the phase transition, the bipartite charge fluctuations discussed in section \ref{sec:bipartite}.

\section{Bipartite charge fluctuations of a critical $c=m$ bosonic model}\label{app:bipartite-c=m}

We extend in this Appendix the results obtained in Section \ref{subsec:bipartite-c=1}.

Let us consider this time a n-channel fermionic wire. Let $(\phi_p, \theta_p)_{1\le p \le n}$ be the corresponding modes obtained by bosonization \protect\cite{QP1DGiamarchi2004}. Suppose that the critical model of the wire is characterized by a central charge $c=m$, with $m$ integer, and that we can find $m$ independent real bosonic modes $(\phi_\alpha, \theta_\alpha)_{1\le \alpha \le m}$, linear combination of $(\phi_p, \theta_p)$, whose effective Hamiltonian is free. The $n-m$ other orthogonal modes are of course necessarily gapped. Then, the logarithmic contribution to the bipartite charge fluctuations in any of the p-channel is positive and can be expressed in terms of the Luttinger parameters of the $\alpha$ modes. We present a quick proof for $n=2$ and $m=1$, which is the case we are presently interested in. The generalization, if fastidious, is self-explanatory.\\ Let $(\phi_1, \theta_1)$ and $(\phi_2, \theta_2)$ be the "good" bosonic modes to describe the system, with $(\phi_1, \theta_1)$ free while $(\phi_2, \theta_2)$ is gapped in $\phi_2$. Let $F_{1}(l)$ and $F_{2}(l)$ be their bipartite charge fluctuations.  The typical Hamiltonian, closed to the fixed point, will be given by $H_c=H_1+H_2$, with:
\begin{equation*}
H_1=\frac{v_{F,1}}{2\pi}\int \frac{1}{K_1}(\nabla \phi_1)^2 + K_1(\nabla \theta_1)^2
\end{equation*}
\begin{equation*}
H_2=\frac{v_{F,2}}{2\pi}\int \frac{1}{K_2}(\nabla \phi_2)^2 + K_2(\nabla \theta_2)^2+g\cos(\alpha \phi_2)
\end{equation*}
with $K_2$, $g$ and $\alpha$ such that $g\cos(\alpha \phi_2)$ is a relevant term that is flowing to strong coupling. A standard computation\protect\cite{QP1DGiamarchi2004} gives:

\begin{align*}
\Braket{e^{i\phi_1(l)}e^{-i\phi_1(0)}} &\approx e^{-\frac{K_1}{2} C(l)} \\
\Braket{e^{i\theta_1(l)}e^{-i\theta_1(0)}} &\approx e^{-\frac{K_1^{-1}}{2} C(l)} \\
\Braket{(\phi_2(l)-\phi_2(0))^2} &=O(1)\\
\Braket{e^{i\theta_2(l)}e^{-i\theta_2(0)}} &\approx e^{-\tilde{\Delta} l}
\end{align*}
with $\tilde{\Delta}$ a non-universal quantity corresponding to the gap of the system, and $C$ a function defined by\protect\cite{QP1DGiamarchi2004}:
\begin{equation*}
C(l)=\frac{1}{2} \log\left(\frac{l^2+\alpha^2}{\alpha^2}\right).
\end{equation*} 
$\alpha$ is a short distance cut-off. Other correlators are zero at the fixed RG point. Using the following equality $
\Braket{e^{i\phi(l)}e^{-i\phi(0)}}=e^{-1/2\Braket{(\phi(l)-\phi(0))^2}}
$, valid for Gaussian modes, one can obtain $F_1$ and $F_2$.  Consequently, $F_1(l)$ scale logarithmically with the length of $A$, while $F_2$ is globally constant. 

Now let $(\phi_{a/b}, \theta_{a/b})$ be two bosonic modes whose charge fluctuations $F_{a/b}$ we can actually access. Assume there exists a unitary hermitian transform mapping the "good" modes to those measured. Let $\phi_{a/b}=\sum\limits_i u_{i,{a/b}} \phi_i + v_{i, {a/b}} \theta_i$. 
One can easily express $F_{a/b}$ as a sum of correlators of $\phi_{1/2}$ and $\theta_{1/2}$. We introduce the quantity $J_i(l)=\frac{1}{\pi^2} \Braket{(\theta_i(l)-\theta_i(0))^2}$. $J_i$  is the analogous of $F_i$, replacing the field $\phi$ by its conjugate $\theta$. It corresponds to the bipartite current fluctuations. It has similar properties. As all cross-correlators cancel close to the Renormalization Group fixed point, we obtain
\begin{align*}
F_{a/b}(l) &= \sum\limits_i u_{i, {a/b}}^2 F_i + v_{i, {a/b}}^2 J_i  \\
F_{a/b}(l) &= (u_{1, {a/b}}^2 \frac{K_1}{\pi^2} + v_{1, {a/b}}^2 \frac{K_1^{-1}}{\pi^2}) \log l +v_{2, {a/b}}^2 \tilde{\Delta} l + {\cal O}(1).
\end{align*}
As all present bosonic modes are real, all coefficients must also be real. As the transformations must be invertible, $u_{1, {a/b}}$ and $v_{1, {a/b}}$ cannot be all zeros. 

Then, there is a logarithmic contribution in at least one of the two observed channels and it must be positive.

Instead of considering the field $(\phi_2, \theta_2)$ gapped in $\phi_2$, one can also gap the mode by fixing $\theta_2$. The role of $F_2$ and $J_2$ is then inverted and our conclusion is still valid.

\section{Bipartite charge fluctuations for Kitaev model from the discrete model}\label{app:bipartite-c=1/2}
\subsection{At the critical point}
We detail here the computation of the bipartite charge fluctuations from the lattice model. From Bogoliubov quasi-particles (see Appendix \ref{app:Bogo}), one can recover the exact expression for $F_A$. In the thermodynamic limit, we obtain:
\begin{multline}
F_A(l)=\iint\limits_{[0,2\pi]^2} \frac{dkdq}{16 \pi^2} \frac{\sin((k-q)l/2)^2}{\sin((k-q)/2)^2}\\
(1+\sin\theta_k \sin\theta_{q}-\cos\theta_k \cos\theta_q),
\end{multline}
where $\theta_k$ is the angle defined in Appendix \ref{app:Bogo}. $F_{Fejer}(k-q, l)=\frac{1}{l} \frac{\sin((k-q)l/2)^2}{\sin((k-q)/2)^2}$ is the Fej\'{e}r Kernel. One can therefore compute the leading linear coefficient of $F_A$ by using Fej\'{e}r theorem:
\begin{align}
F_A(l)&= \int\limits_{[0,2\pi]} \frac{dk}{8 \pi} (1+\sin^2 \theta_k -\cos^2 \theta_k) + {\cal O}(l) \label{Eq:FejerApp-2} \\
&=\frac{|\Delta|}{2|\Delta|+2t} l + {\cal O}(l). \label{Eq:FejerApp}
\end{align}
To obtain the sub-dominant logarithm coefficient, a more involved computation is required. For $\Delta = t$, one can proceed to the complete computation. We introduce the three following auxiliary integrals:
\begin{align*}
 F_A^0&=\iint\limits_{[0,2\pi]^2} \frac{dkdq}{16 \pi^2} \frac{\sin((k-q)l/2)^2}{\sin((k-q)/2)^2}\\
F_A^1&=\iint\limits_{[0,2\pi]^2} \frac{dkdq}{16 \pi^2} \frac{\sin((k-q)l/2)^2}{\sin((k-q)/2)^2}
\sin\theta_k \sin\theta_{q}\\
F_A^2&=\iint\limits_{[0,2\pi]^2} \frac{dkdq}{16 \pi^2} \frac{\sin((k-q)l/2)^2}{\sin((k-q)/2)^2}\cos\theta_k \cos\theta_q
 \end{align*} 
 $F_A^0$ can be actually trivially be computed: $F_A^0(l)=\frac{l}{4}$. For the last two, following an usual trick, we define the following additional integrals: $\Delta F_A^i(l)=F_A^i(l+2)-F_A^i(l)$ and $\Delta \Delta F_A^i(l)=\Delta F_A^i(l+1)-\Delta F_A^i(l)$. Using the trigonometric identity:
\begin{multline*}
(l+3)F_{Fejer}(k, l+3)-(l+2)F_{Fejer}(k, l+2)\\
-(l+1)F_{Fejer}(k, l+1)+l F_{Fejer}(k, l)\\
=4 \cos(\frac{k}{2}) \cos(k(l+\frac{3}{2})),
\end{multline*}
one can compute $\Delta \Delta F_A^i(l)$: 
\begin{align*}
\Delta \Delta F_A^1(l)&=\frac{8(8l^4+48l^3+102l^2+90l+29)}{(2l+1)^2(2l+3)^2(2l+5)^2\pi^2}\\
\Delta \Delta F_A^2(l)&=\frac{4(4l^2+12l+13)}{(2l+1)^2(2l+3)^2(2l+5)^2 \pi^2}.
\end{align*}
From there, a careful but simple resummation allows to obtain the final result:
\begin{equation}
F_A(l)= \frac{l}{4}-\frac{1}{2\pi^2}\log(l)-\frac{\gamma_\text{euler}+2\log(2)}{2\pi^2}+{\cal O}(1).
\end{equation}
The logarithmic contribution for the bipartite charge fluctuations are this time negative. Numerical computations of the relevant integrals confirm that the results stand for all $\Delta/t \neq 0$. \\

\subsection{Corrections around $\mu=\pm 2t$}

We are also interested in computing the leading contribution to the bipartite charge fluctuations around the critical line. We obtain the leading coefficient starting from Eq. \ref{Eq:FejerApp-2}.

\begin{equation}
F_A(l)=\int\limits_{[0,2\pi]} \frac{dk}{4 \pi} \sin^2 \theta_k + {\cal O}(l) 
\end{equation}
Defining:
\begin{align*}
A&=\sqrt{\mu^2+4\Delta^2-4t^2}\\
B_\pm&=\sqrt{4t^2- \mu^2-\Delta^2(4t^2+\mu^2)\pm 2 \Delta \mu A},
\end{align*}
after some cumbersome computations, one can obtain the following expression for the linear coefficient:
\begin{widetext}
\begin{multline}
\frac{\Delta}{4 \pi (\Delta^2-t^2) A}\left( 2 \Delta A \pi +\right. 
\left. i B_- \left(\text{Arg}\frac{2t^2-2\Delta^2-\mu- \Delta A}{B-}-\text{Arg}\frac{-2t^2+2\Delta^2+\mu+ \Delta A}{B-}\right)+\right. \\
\left. i B_+ \left(\text{Arg}\frac{-2t^2+2\Delta^2+\mu- \Delta A}{B+}-\text{Arg}\frac{2t^2-2\Delta^2-\mu+ \Delta A}{B+}\right) \right)
\end{multline}
\end{widetext}

These expressions can be simplified. In the topological phase, the coefficient is constant and equal to its value on the critical line: 
\begin{equation}
\frac{|\Delta |}{2 |\Delta |+2t}
\end{equation}.
Outside the topological phase, assuming $\mu<-2t$, the linear coefficient can be rewritten in the following way:
\begin{equation}
\left\lbrace \begin{array}{ll}
\frac{\Delta}{4 \pi (\Delta^2-t^2) A}\left( 2 \Delta A \pi +i \pi (B_+-B_-) \right) & \text{if $\Delta/t<1$} \\
\frac{t^2}{\mu^2} & \text{if $\Delta/t=1$} \\
\frac{\Delta}{4 \pi (\Delta^2-t^2) A}\left( 2 \Delta A \pi +i \pi (B_+ +B_-) \right)& \text{if $\Delta/t>1$}
\end{array} \right.
\end{equation}

\end{appendix}
\newpage


\bibliography{Papier_v7}

\end{document}